\documentclass[]{elsart}
\usepackage{graphicx}
\usepackage{natbib,rotfloat,count1to,relsize}
\usepackage{txfonts}
\begin{document}

%
\def\aj{AJ }%
\def\araa{ARA\&A }%
\def\apj{ApJ }%
\def\apjl{ApJ }%
\def\apjs{ApJS }%
\def\ao{Appl.~Opt. }%
\def\apss{Ap\&SS }%
\def\aap{A\&A }%
\def\aapr{A\&A~Rev. }%
\def\aaps{A\&AS }%
\def\azh{AZh }%
\def\baas{BAAS }%
\def\jrasc{JRASC }%
\def\memras{MmRAS }%
\def\mnras{MNRAS }%
\def\pra{Phys.~Rev.~A }%
\def\prb{Phys.~Rev.~B }%
\def\prc{Phys.~Rev.~C }%
\def\prd{Phys.~Rev.~D }%
\def\pre{Phys.~Rev.~E }%
\def\prl{Phys.~Rev.~Lett. }%
\def\pasp{PASP }%
\def\pasj{PASJ }%
\def\qjras{QJRAS }%
\def\skytel{S\&T }%
\def\solphys{Sol.~Phys. }%
\def\sovast{Soviet~Ast. }%
\def\ssr{Space~Sci.~Rev. }%
\def\zap{ZAp }%
\def\nat{Nature }%
\def\iaucirc{IAU~Circ. }%
\def\aplett{Astrophys.~Lett. }%
\def\apspr{Astrophys.~Space~Phys.~Res. }%
\def\bain{Bull.~Astron.~Inst.~Netherlands }%
\def\fcp{Fund.~Cosmic~Phys. }%
\def\gca{Geochim.~Cosmochim.~Acta }%
\def\grl{Geophys.~Res.~Lett. }%
\def\jcp{J.~Chem.~Phys. }%
\def\jgr{J.~Geophys.~Res. }%
\def\jqsrt{J.~Quant.~Spec.~Radiat.~Transf. }%
\def\memsai{Mem.~Soc.~Astron.~Italiana }%
\def\nphysa{Nucl.~Phys.~A }%
\def\physrep{Phys.~Rep. }%
\def\physscr{Phys.~Scr }%
\def\planss{Planet.~Space~Sci. }%
\def\procspie{Proc.~SPIE }%
\let\astap=\aap
\let\apjlett=\apjl
\let\apjsupp=\apjs
\let\applopt=\ao

\begin{frontmatter}

\title{The composition and size distribution of the dust in the coma of comet Hale-Bopp}

\author[Pannekoek]{M. Min,}
\ead{mmin@science.uva.nl}
\author[Pannekoek]{J.~W. Hovenier,} 
\author[Pannekoek]{A. de Koter,} 
\author[Pannekoek,Leuven]{L.~B.~F.~M. Waters,} 
\author[Pannekoek]{C. Dominik}

\address[Pannekoek]{Astronomical institute Anton Pannekoek, University of Amsterdam, Kruislaan 403, 1098 SJ  Amsterdam, The Netherlands}
\address[Leuven]{Instituut voor Sterrenkunde, Katholieke Universiteit Leuven, Celestijnenlaan 200B, 3001 Heverlee, Belgium}

\begin{abstract}
We discuss the composition and size distribution of the dust in the
coma of comet Hale-Bopp. We do this using a model fit for the infrared
emission measured by the infrared space observatory (ISO) and the
measured degree of linear polarization of scattered light at various
phase angles and wavelengths. The effects of particle shape on the
modeled optical properties of the dust grains are taken into account.
Both the short wavelength (7-44\,$\mu$m) and the long wavelength
(44-120\,$\mu$m) infrared spectrum are fitted using the same dust
parameters, as well as the degree of linear polarization at $12$
different wavelengths in the optical to near infrared domains. We
constrain our fit by forcing the abundances of the major rock forming
chemical elements to be equal to those observed in meteorites. The
infrared spectrum at long wavelengths reveals that large grains are
needed in order to fit the spectral slope. The size and shape
distribution we employ allows us to estimate the sizes of the
crystalline silicates. The ratios of the strength of various forsterite
features show that the crystalline silicate grains in Hale-Bopp must be
submicron sized. On the basis of our analysis the presence of large
crystalline silicate grains in the coma can be excluded. Because of this lack of large crystalline grains combined with the fact that we do need large amorphous grains to fit the emission spectrum at long wavelengths, we need only approximately 4\% of crystalline silicates by mass (forsterite and
enstatite) to reproduce the observed spectral features.
After correcting for possible hidden crystalline material included in large
amorphous grains, our best estimate of the total mass fraction of
crystalline material is $\sim$7.5\%, which is significantly lower than
deduced in previous studies in which the typical derived crystallinity
is $\sim$${20-30}$\%. The implications of this low abundance of
crystalline material on the possible origin and evolution of the comet
are discussed. We conclude that the crystallinity we observe in
Hale-Bopp is consistent with the production of crystalline silicates in
the inner solar system by thermal annealing and subsequent radial
mixing to the comet forming region ($\sim$30\,AU).
\end{abstract}


\end{frontmatter}
%

\section{Introduction}

Comet Hale-Bopp is undoubtedly the best studied long period comet
in the solar system. The unusual brightness of this comet allowed for
its discovery when it was still at 7\,AU distance from the Earth, and a
long term monitoring of the dust activity after perihelion up to almost
13\,AU \citep{2003A&A...403..313W}. The infrared spectrum taken by the
Infrared Space Observatory (ISO) provided a unique opportunity to study
the composition of cometary dust \citep{1997Sci...275.1904C,
1998A&A...339L...9L}. The strong resonances visible in this spectrum
were attributed to the presence
of crystalline silicates.

The relatively sharp $9.8\,\mu$m feature in the ISO spectrum together
with observations of an extremely high degree of linear polarization
lead to the conclusion that the dust in the coma of Hale-Bopp has an
overabundance of submicron sized grains \citep[see e.g.][and references
therein]{1999EM&P...79..247H, 2001ApJ...549..635M}.

The composition of the dust in Hale-Bopp has been modeled frequently in
the literature applying various approaches \citep[see
e.g][]{1998A&A...339L...9L, 1998ApJ...498L..83L, 1999P&SS...47..773B,
1999ApJ...517.1034W, 2002ApJ...580..579H, 2003A&A...401..577B,
2003ApJ...595..522M}. The interpretation of the results obtained from
the observations lead to discussions in the literature on the origin
and evolution of cometary dust \citep[for a review see][]{2002EM&P...89..247W}. Cometary dust is believed to be the most
primitive material present in the solar system. The high crystallinity
found by most studies of the infrared emission spectrum presents a
problem for this scenario. Since the dust that entered the proto-solar
nebula from the interstellar medium is almost completely amorphous
\citep{2001ApJ...550L.213L, 2004ApJ...609..826K}, the crystalline
silicates in Hale-Bopp have to be products of processing in the early
solar system. Crystalline silicates can only be produced in high
temperature environments ($\gtrsim 1000\,$K), so close to the Sun,
while comets form in regions where the temperatures are low enough for
water ice to exist ($\lesssim 160\,$K). Thus, the crystalline silicates
have to be transported outwards to the regions where the comets form,
or the dust temperature must be locally increased due to for example
lightning or shock annealing
\citep{2002ApJ...565L.109H,1998A&A...331..121P, 2000Icar..143...87D}.

It is clear that in order to constrain the models explaining the
crystalline silicates in cometary material, accurate knowledge of the
abundances of the various components in the dust is required. However,
the large differences in the derived crystalline silicate abundances
published in the literature so far make progress in this area
difficult. All studies agree that the abundance of sub-micron sized
crystalline silicates is very high (although again with considerable
spread), but the question remains open what the crystallinity of the
larger grain population is, where most of the mass resides. This is
because modeling the optical properties of large, irregularly shaped
particles composed of material with high refractive index is
difficult.

In this study, we combine for the first time both the available
observations of the thermal emission as obtained by ISO
\citep{1997Sci...275.1904C, 1998A&A...339L...9L} and the degree of
linear polarization taken from \citet{1998A&AS..129..489G, 1999EM&P...78..373J, 1999EM&P...78..353H, 2000Icar..145..203M} and \citet{2000Icar..143..338J}, with a single dust model. In contrast with previous studies we calculate the optical
properties of the dust grains using a method where we can take into
account both size and shape effects. With this method, we obtain
\emph{information on the composition and mass of the large grain
component}. Also we constrain the chemical abundances as known from
studies of meteorites. Thus, we are able to better constrain the dust
composition and size distribution in the model than previous studies.
As we will show, the size distributions of crystalline and
amorphous silicates are very different, and we are able to rule out the
presence of substantial amounts of large (1-10 $\mu$m) crystalline
silicates. This results in a much lower overall crystallinity than most
other studies found.

In section \ref{sec:method} we explain the fitting procedure and the
dust model we employ. The results of the best fit model are presented
in section \ref{sec:results}. The implications of these results are
discussed in section \ref{sec:discussion}.

\section{Method}
\label{sec:method}

In this section we will outline the method used to interpret the
observations of the thermal infrared emission and the degree of linear
polarization of comet Hale-Bopp in terms of the properties of the
particles in the coma. The method we use has a few important
characteristics:
\begin{itemize}
\item Both size and shape effects of the dust grains are taken into account
when calculating the optical properties, i.e. the absorption (emission) cross
sections and the degree of linear polarization of scattered light. In contrast
with previous studies, our shape distribution enables us to use the same particle
shapes for all dust components and grain sizes.
This implies that we can, for the first time, include the effects of
grain size on the optical properties of crystalline silicate grains.
\item The parameters of the size distributions and the abundances of the various
components are determined using an 'objective' least squares fitting routine.
\item We require the best fit model to fulfill the constraint that the dust in comet
Hale-Bopp has the same abundances of the chemical elements of the major
solid state materials as observed in interplanetary dust particles and
meteorites.
\end{itemize}

\subsection{Size and shape of the dust grains}

The optical properties at a particular wavelength, i.e. the absorption
and scattering cross sections as well as the scattering matrix, of a
dust grain are determined by its size, shape, orientation, structure
and chemical composition. Throughout this paper the size of a dust
grain is defined as the radius of a volume equivalent sphere, $r$. We
use for the size distribution a powerlaw given by
\begin{equation}
\label{eq:power distr} n(r)=\left\{ \begin{array}{ll}
C\cdot r^{\,\beta}, & \qquad r_\mathrm{min}\leq r\leq r_\mathrm{max}, \\
0, & \qquad \textrm{elsewhere}. \end{array}\right.
\end{equation}
Here $n(r)dr$ is the number of dust particles of an ensemble with sizes
between $r$ and $r+dr$; $r_{\mathrm{min}}$ and $r_{\mathrm{max}}$ are
the minimum and maximum grain size, respectively, in the size
distribution; $\beta$ is the index of the powerlaw, and $C$ is a
normalization constant. We choose $C$ so that
\begin{equation}
\int_{r_\mathrm{min}}^{r_\mathrm{max}}\frac{4}{3}\pi \rho r^3n(r)dr=1,
\end{equation}
where $\rho$ is the density of the material considered. Hence, $n(r)dr$
is the number of dust particles with sizes between $r$ and $r+dr$ per
unit mass. The effective radius, $r_\mathrm{eff}$, and the effective
variance, $v_\mathrm{eff}$, are often used to characterize a size
distribution. They are defined by \citep{MishTravis2002}
\begin{eqnarray}
\label{eq:reff}
r_\mathrm{eff}&=&\frac{1}{\left<G\right>}\int_{r_\mathrm{min}}^{r_\mathrm{max}}\pi r^3n(r)dr,\\
\label{eq:veff}
v_\mathrm{eff}&=&\frac{1}{\left<G\right>r_\mathrm{eff}^2}\int_{r_\mathrm{min}}^{r_\mathrm{max}}\pi
r^2(r-r_\mathrm{eff})^2n(r)dr,
\end{eqnarray}
where
\begin{equation}
\left<G\right>=\int_{r_\mathrm{min}}^{r_\mathrm{max}}\pi r^2n(r)dr,
\end{equation}
is the average geometrical shadow of the dust grains. The
$r_\mathrm{eff}$ is simply the surface-area weighted mean whereas
$v_\mathrm{eff}$ is the surface-area weighted variance of the
distribution.

The shape of the dust grains is an important parameter determining
their optical properties. Pictures of interplanetary dust particles
(IDPs) show very complex shapes and structures \citep{DustCatalog}. To
model these complex shapes in detail requires much computing time,
which limits the number of other particle parameters that can be
studied. In \citet{2003A&A...404...35M} it is shown that when the
absorption cross sections of small particles are considered their
shapes can be roughly divided into two categories. The first category
contains homogeneous spherical particles, whereas the second category
contains all other particle shapes. Effects of the specific particle
shape distribution on the absorption properties in the second category
are present, but small compared to the differences with homogeneous
spheres. This implies that the exact shape of the particles is, in a
first approximation, not important as long as we brake the homogeneity
or perfect spherical symmetry of a homogeneous sphere. This approach is
an example of the statistical approach \citep[see
e.g.][]{2002JQSRT..74..167K,2004JQSRT..85..231K}. In this approach the
absorption properties of an ensemble of irregular particles are
simulated by the average absorption properties of an ensemble of
particles with simple shapes.

In this paper we use a distribution of hollow spherical particles
(DHS). In this distribution a uniform average is taken over the
fraction, $f$, of the total volume occupied by the central vacuum
inclusion in the range $0\leq f< f_{\mathrm{max}}$, where
$f_\mathrm{max}\leq 1$. All particles in this distribution have the
same material volume, so that the particles with higher values of $f$
have larger outer radii. For details regarding this distribution we
refer to \citet{2003A&A...404...35M}. In order to reproduce the
wavelength positions of the spectral features of crystalline silicate
grains, we have to choose $f_\mathrm{max}=1$
\citep{2003A&A...404...35M}. Numerical computations for particles with
$f=1$ are not possible since these would have an infinitely large outer
radius. For particles much smaller than the wavelength, integrating the
optical properties up to $f=1$ can be done analytically
\citep{2003A&A...404...35M}. For larger particles numerical
computations are necessary. We then choose $f_\mathrm{max}$ large
enough to reach convergence to the values for $f_\mathrm{max}=1$. For
most cases we considered, it suffices to integrate up to $f=0.98$. The
optical properties of hollow spherical particles are calculated using a
simple extension of Mie theory \citep{Aden1951}. The calculations were
done using the layered sphere code for which the basic ideas are
explained in \citet{1981ApOpt..20.3657T}. We would like to stress that
we do not allege that hollow spheres are a good approximation for the
real shape of cometary dust grains. Rather, the properties of the
particles in the DHS determining the absorption and polarization
behavior of the entire ensemble represent in a statistical way those of
an ensemble of realistically shaped dust grains.

One of the advantages of the DHS is that the scattering and absorption
properties can be calculated easily for almost all grain sizes and
wavelengths. This means that we can calculate the absorption cross
sections as well as the degree of linear polarization for incident
unpolarized light as a function of scattering angle, which enables us
to fit simultaneously the observed infrared emission spectrum and the
degree of linear polarization using the same dust parameters. In
\citet{2003A&A...404...35M} it is shown that good agreement between
calculations for the distribution of hollow spheres and measurements of
the mass absorption coefficients of small forsterite grains as a
function of wavelength can be obtained. Also, as is shown by \citet{MinHollow}, the degree of linear polarization as calculated using
the DHS is in good agreement with laboratory measurements of
irregularly shaped quartz particles.

A cometary dust grain consists most likely of a mixture of various
components such as, for example, olivine, carbon and ironsulfide. In
using the model described above we assume that the optical properties
of such a mixed particle can be represented by the average properties
of homogeneous particles for each of the separate components. 
The validity of this approach for core-mantle grains is studied by \citet{2002MNRAS.334..840L} and shown to be dependent on the shape of the dust grains. In general, whether it is a valid assumption depends on the compactness of the composite particle.
When the grains are very 'fluffy' aggregates composed of
homogeneous monomers, the grain components will approximately interact
with the incoming light as if they were separate. However, in a more
compact structure, the effects of interaction between the separate
components can become visible.

Throughout the paper we will first average over particle shape ($f$)
which will be denoted by $\left<..\right>$. The averaging over particle
shape and size will be denoted by $\left<\left<..\right>\right>$.

\subsection{Thermal emission}

To compute the radiation emitted by an ensemble of monodisperse dust
grains in random orientation as a function of wavelength we need two
ingredients, {\it i)} the orientation averaged absorption cross section
as a function of wavelength, and {\it ii)} the temperature of the dust
grains. When both are known we can calculate the observed thermal
radiation of a dust grain averaged over all orientations as follows
\begin{equation}
\label{eq:flux} \mathcal{F}(\lambda)=
\frac{C_{\mathrm{abs}}(\lambda)B_{\lambda}(T)}{D^2}.
\end{equation}
Here $\mathcal{F}(\lambda)$ is the flux density at distance $D$, $\lambda$ is
the wavelength of radiation, $T$ is the temperature of the dust grain,
$B_{\lambda}(T)$ is the Planck function, and
$C_{\mathrm{abs}}(\lambda)$ is the orientation averaged absorption
cross section of the dust grains at wavelength $\lambda$. 
The thermal radiation of grains with the temperatures we consider is
mainly emitted in the infrared part of the spectrum. At these
wavelengths the effect of scattering of solar radiation on the total
observed flux is negligible.

The temperature of the dust grains is calculated by assuming thermal
equilibrium, i.e. the energy absorbed is equal to the energy emitted.
Note that dust grains with equal size $r$ and composition but different
shapes and/or orientations can have different equilibrium temperatures.
However, we assume that all dust grains with the same volume and
composition have the same temperature determined by the shape and
orientation averaged absorption properties (as is usually done).

The coma of Hale-Bopp is optically thin. Therefore, we only need to
consider direct illumination of the coma grains by the Sun and we may
ignore the diffuse radiation field caused by, e.g., scattering of solar
light by the coma material. We assume the Sun to radiate like a black
body with a temperature of $T=5777$\,K.

Since the absorption cross section, and, therefore, the temperature,
depends on the size and chemical composition of the particles we
determine the temperature for each different dust particle size and
material separately. Using the size distribution given by
Eq.~(\ref{eq:power distr}) the flux per unit mass from dust component $j$ is given by
\begin{equation}
\left<\left<\mathcal{F}(\lambda)\right>\right>_j=\frac{1}{D^2}\,\int_{r_{\mathrm{min},j}}^{r_{\mathrm{max},j}}n_j(r)
\left<C_{\mathrm{abs},j}(\lambda,r)\right>B_{\lambda}(T_j(r))~dr.
\label{eq:flux per material}
\end{equation}
In this equation $n_j(r)$ is the size distribution of component $j$
with minimum and maximum radii $r_{\mathrm{min},j}$ and
$r_{\mathrm{max},j}$; $\left<C_{\mathrm{abs},j}(\lambda,r)\right>$ is
the orientation and shape averaged absorption cross section of an
ensemble of dust grains of material $j$ with size $r$, and $T_j(r)$ is
the temperature of the dust grains in this ensemble. Since the coma of
comet Hale-Bopp is optically thin, the total flux averaged over
particle size, shape, orientation and composition is simply the sum
over the various components
\begin{equation}
\mathcal{F}_\mathrm{model}(\lambda)=\sum_j M_j
\left<\left<\mathcal{F}(\lambda)\right>\right>_j, \label{eq:total flux}
\end{equation}
where $M_j$ is the total mass of dust component $j$.

\subsection{Degree of linear polarization of scattered light}

In the visible part of the spectrum the radiation from the cometary
halo is dominated by sunlight scattered once by dust grains. The
intensity and polarization of the scattered light depend on the angle
of scattering and the wavelength. For a comet it is possible to obtain
measurements of the degree of linear polarization for various
scattering angles by observing the comet at various moments during its
orbit around the Sun. All information on the angular dependence of the
scattering behavior of an optically thin ensemble of dust grains is
contained in its $4\times 4$ scattering matrix. When the size and shape
distributions and the abundances of all dust species are known, we can
calculate the average $4\times 4$ scattering matrix of the ensemble if
enough data of the refractive index is available. From this matrix we
can obtain the degree of linear polarization for incident unpolarized
light of dust component $j$
\begin{equation}
\left<\left<P(\alpha)\right>\right>_j=-\frac{\left<\left<F_{21}(\alpha)\right>\right>_j}{\left<\left<F_{11}(\alpha)\right>\right>_j},
\end{equation}
where $\alpha$ is the phase angle and
$\left<\left<F_{nk}\right>\right>_j$ is the $n,k$th element of the
scattering matrix averaged over size and shape of the dust grains 
\citep[for details see][]{vandeHulst}. It should be noted
that for calculations and measurements presented in the literature
often the scattering angle, $\theta$, is used instead of the phase
angle, $\alpha$. For comets it is more convenient to use the phase
angle. Since multiple scattering can be neglected for comets we have
$\alpha=180^\circ-\theta$.

In order to calculate the degree of linear polarization of an ensemble
of particles not only averaged over particle size and shape
distributions but also over dust materials, we have to average the
matrix elements $\left<\left<F_{nk}\right>\right>_j$. Thus the average
polarization is given by
\begin{equation}
\label{eq:polarization model} P_{\mathrm{model}}(\alpha)=-\frac{\sum_j
M_j \left<\left<F_{21}(\alpha)\right>\right>_j}{\sum_j M_j
\left<\left<F_{11}(\alpha)\right>\right>_j}.
\end{equation}

\subsection{Least squares fitting procedure}
\label{sec:fitting procedure}

To make a fit to the observations of Hale-Bopp we need to fine-tune the
free parameters in the model in such a way that we minimize the
differences between the results of the model computations and the
observations in a well defined way. The model we constructed has $4N_d$
free parameters, where $N_d$ is the number of dust species we include
in the fitting procedure. By choosing the index of the powerlaw,
$\beta$, the same for all dust species this reduces to $3N_d+1$ free
parameters, namely $\left(\{
r_{\mathrm{min},j},r_{\mathrm{max},j},M_j,j=1..N_d\},\beta\right)$. The
error on the infrared emission spectrum is defined as
\begin{equation}
\label{eq:chisquare spec} \chi^2_{\mathrm{spec}}=\sum_{i=1}^{N_\lambda}
\left|\frac{\mathcal{F}_\mathrm{model}(\lambda_i)-\mathcal{F}_{\mathrm{obs}}(\lambda_i)}{\sigma_\mathcal{F}(\lambda_i)}\right|^2.
\end{equation}
In this equation $\lambda_i$ ($i=1..N_\lambda$) is a chosen wavelength
grid; $\mathcal{F}_{\mathrm{obs}}(\lambda_i)$ is the observational
value of the flux at wavelength $\lambda_i$, and
$\sigma_\mathcal{F}(\lambda_i)$ is the error of the observed flux at
wavelength $\lambda_i$. In order to estimate
$\sigma_\mathcal{F}(\lambda_i)$ we assume that the error in the
spectral observations is proportional to the square root of the
observed flux. The position of the minimum value of
$\chi^2_\mathrm{spec}$ is independent of the absolute value of the
error. Note that, since we do not have the values of the absolute
errors of the spectral measurements, the $\chi^2_\mathrm{spec}$ is not
equal to the reduced $\chi^2_\mathrm{spec}$ and cannot be interpreted
as the statistical goodness of fit.

The error of the degree of linear polarization is defined as
\begin{equation}
\label{eq:chisquare pol}
\chi^2_{\mathrm{pol}}=\sum_j^{N'_\lambda}\sum_i^{N_\alpha}\left|\frac{P_\mathrm{obs}(\alpha_i,\lambda_j)-P_\mathrm{model}(\alpha_i,\lambda_j)}{\sigma_P(\alpha_i,\lambda_j)}\right|^2.
\end{equation}
In this equation $P_\mathrm{obs}(\alpha_i,\lambda_j)$ is the degree of
linear polarization observed at phase angle $\alpha_i$
($i=1..N_\alpha$) and wavelength $\lambda_j$ ($j=1..N'_\lambda$), while
$\sigma_P(\alpha_i,\lambda_j)$ is the error in the observed
polarization at phase angle $\alpha_i$ and wavelength $\lambda_j$. Note
that the number of phase angles at which observations are available may
vary with wavelength.

The most straightforward way to define the total $\chi^2$ is to take
the sum of $\chi^2_\mathrm{spec}$ and $\chi^2_\mathrm{pol}$. However,
since we have many more measurements of the infrared flux than we have
observations of the degree of linear polarization and we are unable to
compute the reduced $\chi^2_\mathrm{spec}$, this would lead to a
stronger weight of the spectral measurements than the polarization
observations. Therefore, we chose to minimize
\begin{equation}
\label{eq:chisquare}
\chi^2=\chi^2_{\mathrm{spec}}\cdot\chi^2_{\mathrm{pol}}.
\end{equation}

The minimization of $\chi^2$ as defined by Eq.~(\ref{eq:chisquare}) is
done using a combination of two methods. Since the total spectrum is a
linear combination of the separate spectra for the different components
(see Eq.~\ref{eq:total flux}), we are able to separate the fitting
problem into a non-linear and a linear part. The non-linear part
consists of minimizing for all $r_{\mathrm{min},j}$,
$r_{\mathrm{max},j}$ and $\beta$, and the linear part does the
minimization for all $M_j$.

For the non-linear part of the minimization we use a genetic
optimization algorithm called \textsc{pikaia}
\citep{1995ApJS..101..309C}. This algorithm tries to find a global
maximum of an arbitrary function in a large parameter space by using
concepts from evolution theory. Every set of parameters $\left(\{
r_{\mathrm{min},j},r_{\mathrm{max},j},j=1..N_d\},\beta\right)$ is
called an individual, the parameters are the 'genes'. The procedure
starts with a randomly initialized population of $N_\mathrm{pop}$
individuals and calculates the $\chi^2$ for each individual. The
individuals with the highest values of $1/\chi^2$ (lowest values of
$\chi^2$) are then given the best chance to 'reproduce' into the next
generation of individuals. This reproduction is done by mixing the
genes of two individuals into a new individual. When this procedure is
repeated through several generations, the individuals with low values
of $1/\chi^2$ will die out and only individuals with high values of
$1/\chi^2$ will survive. In the end (after $N_\mathrm{gen}$
generations) the best individual, which represents the best fit
parameters, will survive. To ensure convergence also mutations --
random variations on the parameters -- are included in the algorithm.
An extensive description can be found in \citet{1995ApJS..101..309C}.
Although the algorithm is not very fast -- many models need to be
calculated -- it is very robust in the sense that it will (almost)
always find the global maximum, whereas other optimization codes
frequently end up with a local maximum.

For every individual in the genetic algorithm we determine the best
values for the $M_j$ by minimizing the $\chi^2_\mathrm{spec}$ using a
linear least squares fitting procedure. Linear least squares fitting
amounts to solving an overdetermined matrix equation in a least squares
sense. To ensure that all $M_j$ are positive we need a robust linear
least squares fitting procedure with extra linear equality and
inequality constraints. We use the subroutine \textsc{dlsei} from the
\textsc{slatec} library\footnote{The \textsc{slatec} library is
publicly available for download at: http://www.netlib.org/slatec/}. The
inequality constraints are used to ensure that $M_j>0$. The equality
constraints are used to constrain the chemical abundances as will be
explained in section \ref{sec:abun constraints}. Using the $M_j$ thus
derived, we calculate $\chi^2_\mathrm{pol}$ and $\chi^2$. Although in
this way we might miss the absolute minimum, we can use this approach
since the infrared spectrum is more sensitive to the exact dust
composition than the degree of linear polarization is, which is more
sensitive to the size distribution.

In order to estimate the errors on the derived abundances of the dust
species we consider all individuals from all generations with a value
of $\chi^2$ smaller than $1.1$ times the minimum value of $\chi^2$. The
value $1.1$ was chosen such that all fits within this range are still
in reasonable agreement with the observations.

\begin{figure*}[!t]
\resizebox{\hsize}{!}{\includegraphics{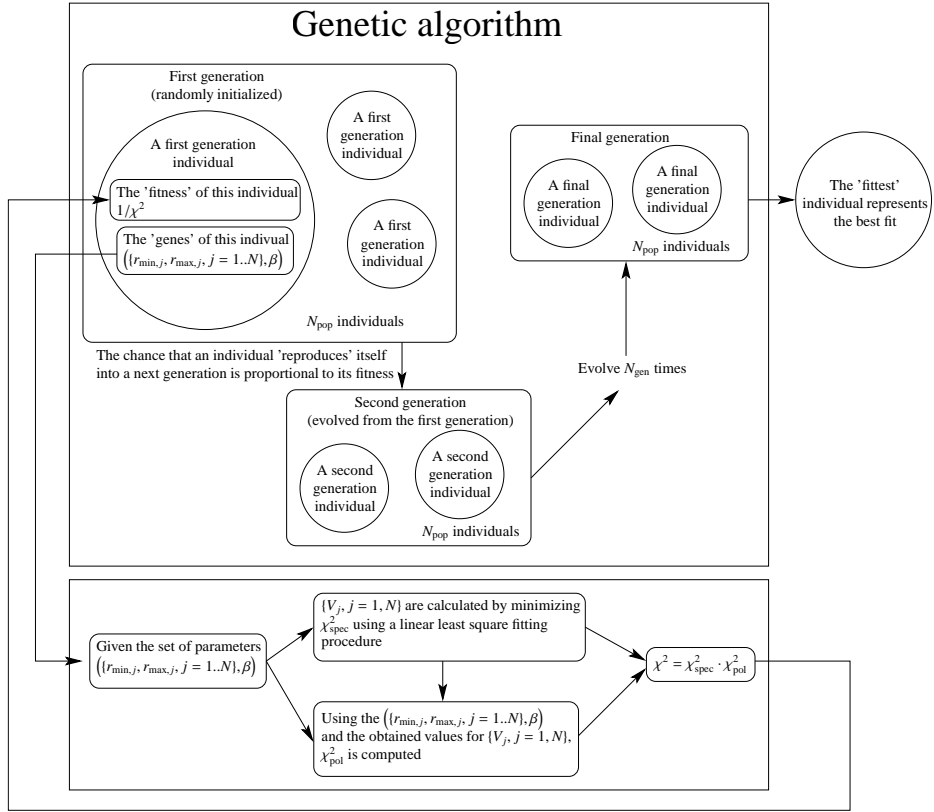}} \caption{A schematic
representation of the fitting procedure.} \label{fig:Diagram}
\end{figure*}

The entire fit procedure is schematically outlined in
Fig.~\ref{fig:Diagram}.

\subsection{Dust components}

In this section we will discuss the dust components we chose for our
model computations. Based on previous studies and on the composition of
IDPs we make the following selection of components.
\begin{description}
\item[$\bullet$ Amorphous Carbon] (C). We assume that most of the Carbon atoms
that are in the solid state phase will be present as amorphous carbon. The
emission spectrum of amorphous carbon gives a smooth continuum contribution
without clear spectral structure. From the \emph{in situ} measurements by the
Vega spacecraft when it encountered the coma of comet Halley, we know that
approximately half of the total available C is present in the solid state
phase \citep{1987A&A...187..859G}. Carbon probably acts as a matrix in which
the other materials are embedded. The refractive index as a function of
wavelength is taken from \citet{1993A&A...279..577P}.
\item[$\bullet$ Amorphous Olivine] (Mg$_{2x}$Fe$_{2-2x}$SiO$_4$). Amorphous
olivine is one of the most abundant dust species in circumstellar and
interstellar material \citep[see e.g.][]{1999A&A...350..163M, 2001A&A...375..950B,
2001ApJ...550L.213L, 2004ApJS..154..443F, 2004ApJ...609..826K}.
The silicate dust in the diffuse interstellar medium is dominated
by amorphous olivine \citep{2004ApJ...609..826K}. The emission by
small amorphous olivine grains shows broad spectral features at
$10$ and $20\,\mu$m. Both are detected in the spectrum of Hale-Bopp. T
he data for the refractive indices are taken from \citet{1995A&A...300..503D}.
\item[$\bullet$ Amorphous Pyroxene] (Mg$_{x}$Fe$_{1-x}$SiO$_3$).
Amorphous pyroxene is spectroscopically hard to distinguish from
amorphous olivine. Pyroxene is often found in IDPs. The emission by
small amorphous pyroxene grains shows a spectral structure which is
similar to that of amorphous olivine grains. However, the maximum of
the $10\,\mu$m feature is shifted towards slightly shorter wavelengths
and the shape of the $20\,\mu$m feature is slightly different. The
refractive indices for amorphous pyroxene are taken from
\citet{1995A&A...300..503D}.
\item[$\bullet$ Crystalline Forsterite] (Mg$_2$SiO$_4$). Crystalline
forsterite is the magnesium rich end member of the crystalline olivine
family. Experiments indicate that when amorphous olivine is annealed
under certain conditions, the iron is removed from the lattice
structure and crystalline forsterite is formed. The most important
resonances in the emission spectrum of small crystalline forsterite
grains are located at wavelengths $11.3$, $19.5$, $23.6$ and
$33.6\,\mu$m. These features are observed in the spectra of, for
example, AGB stars and protoplanetary disks \citep[see
e.g.][]{1996A&A...315L.361W, 2001A&A...375..950B} and are clearly
visible in the spectrum of Hale-Bopp \citep{1997Sci...275.1904C}. For
the refractive indices of forsterite we use the data of
\citet{Servoin}.
\item[$\bullet$ Crystalline Enstatite]  (MgSiO$_3$). Crystalline
enstatite is the magnesium rich end member of the crystalline pyroxene
family. Under the right conditions, enstatite can form from a reaction
between forsterite and silica. Also, from IDPs there are indications
that some enstatite grains formed directly from gas phase condensation
\citep{1983Natur.301..473B}. The spectrum of small enstatite grains
shows prominent features at $9.3$, $10.5$, $19.5$, $28$ and
$44.5\,\mu$m. Although we have no direct evidence for the presence of
most of these features, we include enstatite in the fitting procedure
because enstatite features have been reported in the ground based
spectrum of Hale-Bopp when it was at $1.2$\,AU
\citep{1999ApJ...517.1034W}. The data for the refractive indices are
taken from \citet{1998A&A...339..904J}.
\item[$\bullet$ Amorphous Silica] (SiO$_2$). Laboratory measurements
show that when amorphous silicates are annealed to form crystalline
silicates, a certain amount of silica is also produced \citep{2000A&A...364..282F}.
The emission spectrum of small amorphous silica grains shows features
at $9$, $12.5$ and $21\,\mu$m. We take the refractive indices as a
function of wavelength measured by \citet{1960PhRv..121.1324S}.
\item[$\bullet$ Metallic Iron] (Fe). When amorphous silicates are
annealed to form magnesium rich crystalline silicates, the iron is
removed from the lattice. The spectral signature of metallic iron
grains is very smooth but slightly different than that of amorphous
carbon. Although it is not crucial in obtaining a reasonable fit, we
include the possibility of metallic iron grains in the fitting
procedure for completeness. The refractive indices of metallic iron are
taken from \citet{1996A&A...312..511H}.
\item[$\bullet$ Iron Sulfide] (FeS). In IDPs all available sulfur is
present as iron sulfide. The spectral structure of emission from iron
sulfide grains shows a broad feature around $23\,\mu$m. The fact that
this feature is not prominent in the spectrum of Hale-Bopp is probably
due to the fact that the iron sulfide grains are relatively large,
which reduces the spectral structure significantly. We use the
refractive indices measured by \citet{1994ApJ...423L..71B}.
\end{description}

To decrease the number of free parameters, we have chosen equal size
distributions for both amorphous silicate species (amorphous olivine
and amorphous pyroxene) and equal size distributions for both
crystalline silicate species (forsterite and enstatite). Furthermore
$\beta$, the index of the powerlaw, is the same for all materials.

\subsection{Chemical abundance constraints}
\label{sec:abun constraints}

The number of free parameters in the method as described above is very
large. When we apply no extra constraints we encounter a large set of
solutions that all have a more or less equal $\chi^2$, but with totally
different size distributions and material abundances. We also run the
risk of obtaining a best fit solution with very implausible values for
the fit parameters. To avoid these problems we constrain the solution
by requiring that the major elements for solid state materials (Si, Mg,
Fe and S) are all in the solid phase and that their chemical abundances
are the same as those found in meteorites. Furthermore, we take the
abundance of Carbon in the solid phase to be half the solar abundance,
consistent with \emph{in situ} measurements of the dust in the coma of
comet Halley \citep{1987A&A...187..859G}. This gives us four extra
constraints on the model. These extra constraints prove to be
sufficient to obtain consistent results with the fitting procedure.

We constrain the abundances of C, Mg, Fe and S relative to Si. For the Carbon
abundance we take half of the total solar abundance to be in the solid phase,
the remaining Carbon is in the the gas phase. Magnesium, iron, sulfur and
silicon are assumed to be completely in the solid phase. The abundances we use
are taken from \citet{1998SSRv...85..161G} and are summarized in
Table~\ref{tab:chemical abundances}. Note that the sulfur abundance
measured in meteorites is lower than the solar abundance. Thus by taking the
constraints from meteorites for the solid state particles we assume there is
also sulfur in the gas phase. This is consistent with observations \citep{2000A&A...353.1101B, 2000Icar..143..412I}. All
abundance constraints are incorporated as linear equality constraints to the
linear least squares fitting part of the minimization procedure.

\begin{table}[!t]
\begin{center}
\begin{tabular}{|c||c|c|c|}
\hline
Chemical element & Meteorites & Solar & Constraints \\
\hline \hline
C/Si    &   -   &   9.33    &   4.67\\
Mg/Si   &   1.05    &   1.07    &   1.05\\
Fe/Si   &   0.87    &   0.89    &   0.87\\
S/Si    &   0.44    &   0.60    &   0.44\\
\hline
\end{tabular}
\end{center}
\caption{Abundance constraints as applied in the fitting procedure.
These values are taken from \citet{1998SSRv...85..161G}. For Carbon we
take half of the solar abundance, consistent with \emph{in situ}
measurements of dust in the coma of comet Halley
\citep{1987A&A...187..859G}.} \label{tab:chemical abundances}
\end{table}

To use the chemical abundance constraints we have to introduce an extra
free parameter, $x$, the magnesium fraction in the amorphous silicates.
In section \ref{sec:best fit model} we will show that the bulk of the
material consists of olivine, pyroxene, carbon and iron sulfide. Using
this information we can already make a simple but reliable estimate of
the value of $x$ by adopting these four species only and applying the
abundance constraints discussed above (as this implies that we have
four constraints and four unknown parameters the material abundances
are uniquely defined). The results of this simple calculation are
summarized in Table~\ref{tab:abun estimate}, and give $x=0.7$. Note
that here it is not possible to distinguish between amorphous and
crystalline material. The results from the fitting procedure will be
slightly different due to a different composition of crystalline and
amorphous silicates and the fact that in the fitting procedure we also
added silica and metallic iron.

\begin{table}[!t]
\begin{center}
\begin{tabular}{|c||c|c|}
\hline
Dust component & Chemical formula & Abundance (Mass \%)\\
\hline \hline
Carbon  &C                  &24.6\\
Olivine &Mg$_{2x}$Fe$_{2-2x}$SiO$_4$  &33.4\\
Pyroxene    &Mg$_{x}$Fe$_{1-x}$SiO$_3$  &25.0\\
Iron Sulfide&FeS                    &17.0\\
\hline
\end{tabular}
\end{center}
\caption{Abundances as calculated from chemical abundances found in
meteorites. In order to satisfy the chemical abundances given in Table
\ref{tab:chemical abundances}, we have to take $x=0.7$.}
\label{tab:abun estimate}
\end{table}

\section{Results}
\label{sec:results}

\subsection{Observations}

The spectroscopic observations we use are the infrared spectra obtained
by the Short Wavelength Spectrometer (SWS) and the Long Wavelength
Spectrometer (LWS) on board the Infrared Space Observatory (ISO)
\citep{1997Sci...275.1904C, 1998A&A...339L...9L}. These spectra were
taken when the comet was at $2.9$\,AU distance from the Sun and $3$\,AU
distance from the Earth. The SWS and LWS spectra have a small
overlapping wavelength range ($42\,\mu$m$\,<\lambda<45\,\mu$m). Since
the LWS has a larger beam size, it catches emission from a larger part
of the coma, resulting in a higher absolute flux level. We assume that
the properties of the dust causing the emission does not change as a
function of distance from the core of the comet, so we can simply scale
the LWS spectrum to match the SWS spectrum in the overlapping
wavelength region. We have to note here that there are indications that
the size distribution or the compactness of the particles varies
slightly as a function of position in the coma \citep[see e.g.][and
references therein]{Kolokolova2004}. When the particles move away from
the comet nucleus, the particles might fall apart resulting in more
fluffy, or smaller structures. Therefore, one might argue that taking
into account the extended region covered by the LWS might bias our
results towards a slightly higher fraction of small grains. However, we
believe that this effect is only minor since these differences are
largest when considering the region very close to the coma, which is
covered by both the SWS and the LWS \citep{Kolokolova2004}. For the
fitting procedure we used the wavelength range from $7$ to $120\,\mu$m.

For the observational data of the degree of linear polarization we used
the combined measurements from various studies in the optical to near infrared part of
the spectrum. The measurements were taken from
\citet{1998A&AS..129..489G, 1999EM&P...78..373J, 1999EM&P...78..353H, 2000Icar..145..203M} and \citet{2000Icar..143..338J}. 
The observations also provide the errors $\sigma_P$. 
Combining the observations we have polarization data
at twelve different wavelengths for various phase angles. We note that in
order to obtain observations at different phase angles, the comet has
to be observed at different phases during its orbit around the Sun.
Therefore, in order to model all these observations using a single dust
model, we have to assume that the composition and size distribution of
the dust is more or less constant at different phases. Since we
consider the degree of linear polarization of scattered light,
variations in the total dust mass in the coma are not important.

\subsection{Best fit model}
\label{sec:best fit model}

\begin{figure*}[!t]
\resizebox{\hsize}{!}{\includegraphics{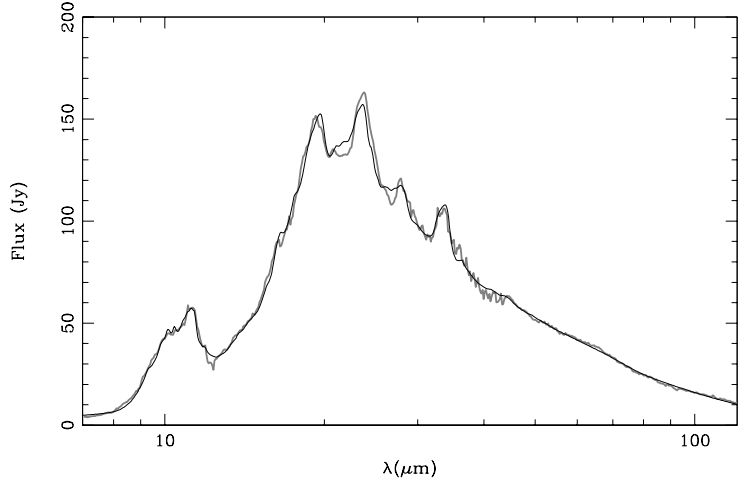}} 
\caption{Infrared
spectral energy distributions of the best fit model (black line)
together with the SWS and the LWS observations (gray line).}
\label{fig:Spectral Fit}
\end{figure*}

\begin{figure*}[!t]
\resizebox{\hsize}{!}{\includegraphics{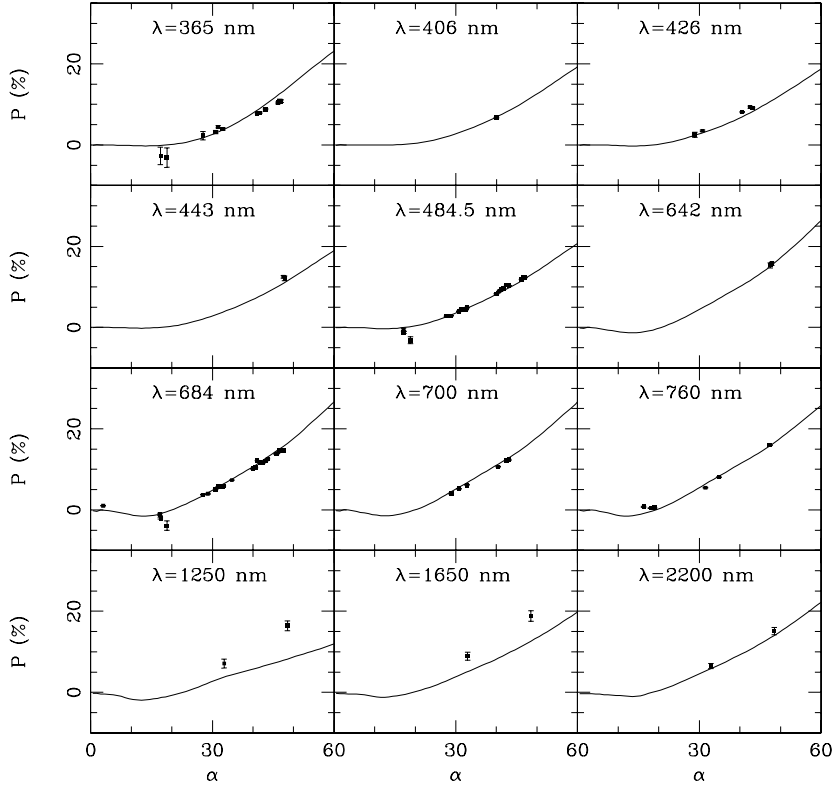}} 
\caption{The degree
of linear polarization from the best fit model as a function of phase
angle, $\alpha$, in degrees (solid curves) based on both the spectral
and the polarization measurements. The measurements of the linear
polarization are indicated by dots and were taken from
\citet{1998A&AS..129..489G, 1999EM&P...78..373J, 1999EM&P...78..353H, 2000Icar..145..203M} and \citet{2000Icar..143..338J}.} 
\label{fig:polarize model}
\end{figure*}

\begin{figure}[!t]
\center{\resizebox{11cm}{!}{\includegraphics{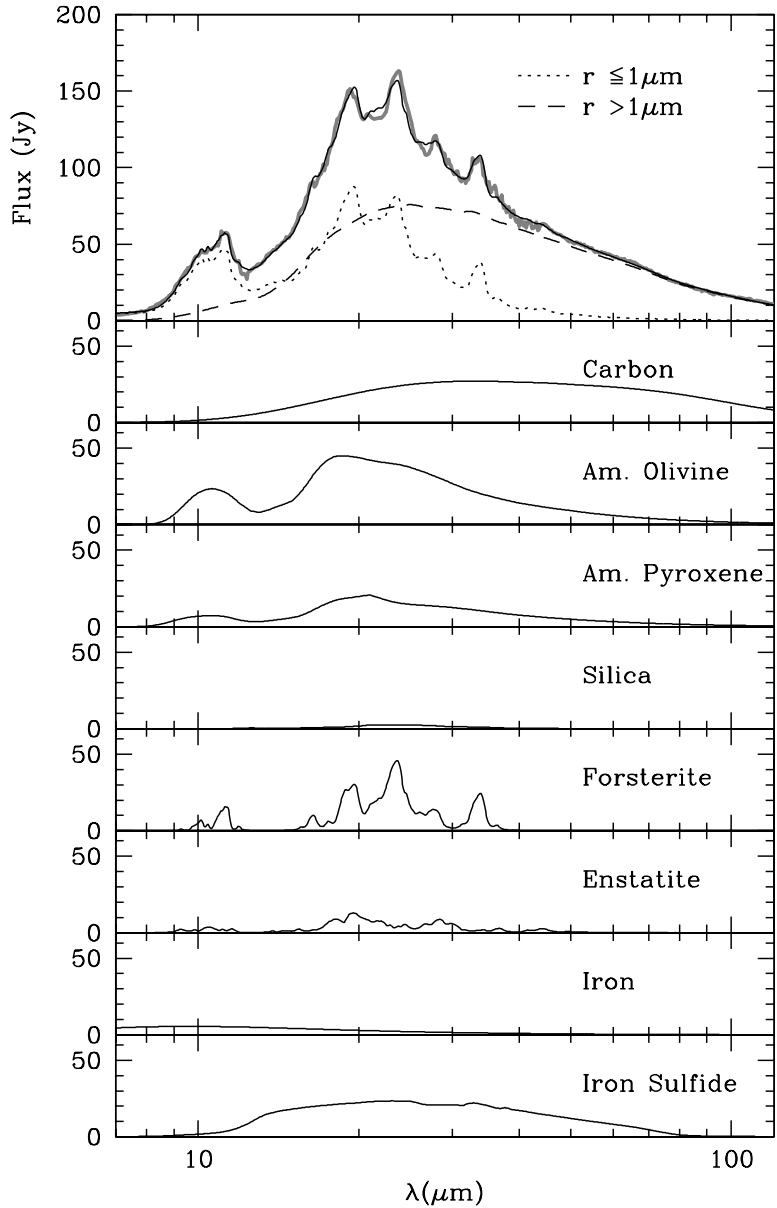}}} 
\caption{The
contributions of all small (dots), and large (dashes) grain components
to the total spectrum (upper panel). The gray line indicates the
observations, the black line indicates the best fit model. Also plotted
are the contributions to the emission spectrum of the various dust
components (lower panels).} 
\label{fig:Fit}
\end{figure}

We minimized $\chi^2$ as defined by Eq.~(\ref{eq:chisquare}) using the
method described in section \ref{sec:fitting procedure}. The
$\lambda_i$ were chosen on a logarithmic grid. Throughout the fitting
procedure we fix the value of $f_\mathrm{max}$ for each dust species.
From the positions of the crystalline silicate resonances we already
know that for these materials we have to choose the most extreme shape
distribution parameters ($f_\mathrm{max}=1$). The other materials are
chosen to have equal values of $f_\mathrm{max}$. For these materials,
we obtain an optimum value $f_\mathrm{max}=0.8$. The parameters are
shown in Table \ref{tab:Fit}. The infrared spectrum corresponding to
the best fit model is shown in Fig.~\ref{fig:Spectral Fit} together
with the measurements. The resulting curves for the degree of linear
polarization as functions of the phase angle, together with the
observations, are shown in Fig.~\ref{fig:polarize model}. The emission
spectra of the separate dust components are plotted in
Fig.~\ref{fig:Fit}.

\begin{sidewaystable}[]
\begin{center}
\begin{tabular}{|l|c|c|c|c@{\quad(}c@{~/~}c@{)~}|c|c|c|c|}
\hline
Material & Chemical Formula & Bulk density & $f_{\mathrm{max}}$ & Volume & min & max & Mass & $r_{\textrm{min}}$ & $r_{\textrm{max}}$ & $\beta$\\
 &  & (g/cm$^3$) &  & \multicolumn{3}{c|}{(\%)}& (\%) & $(\mu$m) & $(\mu$m) &\\
\hline \hline
Amorphous Carbon        &C                  		&1.80&0.8&39.1&39.0&39.3&23.8&7.0 &11.6 &-3.48\\
Amorphous Olivine       &Mg$_{2x}$Fe$_{2-2x}$SiO$_4$	&3.71&0.8&25.7&22.9&32.7&32.3&0.01&92.8 &-3.48\\
Amorphous Pyroxene  	&Mg$_{x}$Fe$_{1-x}$SiO$_3$  	&3.20&0.8&18.3& 7.8&23.5&19.8&0.01&92.8 &-3.48\\
Amorphous Silica        &SiO$_2$                	&2.60&0.8& 2.2& 0.8& 5.6& 1.9&6.5 & 7.5 &-3.48\\
Crystalline Forsterite  &Mg$_2$SiO$_4$          	&3.33&1.0& 2.7& 2.5& 3.0& 3.1&0.05& 0.1 &-3.48\\
Crystalline Enstatite   &MgSiO$_4$              	&2.80&1.0& 1.3& 0.9& 1.7& 1.2&0.05& 0.1 &-3.48\\
Metallic Iron       	&Fe                 		&7.87&0.8& 0.6& 0.3& 1.1& 1.5&0.2 & 0.3 &-3.48\\
Iron Sulfide        	&FeS                    	&4.83&0.8&10.1&10.0&10.1&16.4&0.8 & 4.3 &-3.48\\
\hline
\end{tabular}
\end{center}
\caption{Results for the best fit model. The error estimates are obtained by considering all possible fits with $\chi^2\leq1.1\chi^2_\mathrm{min}$. The total dust mass in the beam of the SWS as found from the fit is $4.6\cdot10^{9}$\,kg.}
\label{tab:Fit}
\end{sidewaystable}

Figs.~\ref{fig:Spectral Fit} and \ref{fig:polarize model} show that the
observations of both the infrared spectrum and the degree of linear
polarization can be reproduced remarkably well using the same dust
model. In other words, we do not need quite different models for the infrared and optical parts of the spectrum. 
The differences between the observed and predicted infrared
spectrum are most probably mainly caused by uncertainties in the
refractive index data, and the assumptions on which the model is based.
For the degree of linear polarization we notice that the negative
polarization branch at $\lambda=365$\,nm (and to a smaller extend at $484.5$ and $684$\,nm) at small phase angles is not
reproduced satisfactorily by the model. This is most likely due to the
spherical symmetry of the shapes we employed. In addition a discrepancy between the model and the observations occurs at $\lambda=1250$ and $1650$\,nm, which is possibly also connected to the adopted spherical particle shapes. The scattered light is
dominated by the contributions from the silicate (olivine and pyroxene)
and ironsulfide grains. The contribution to the scattered light from
carbon grains is negligible. Although the abundance of silicates is
much higher than that of ironsulfide, the scattering caused by
ironsulfide grains is comparable to that of the silicates due to its
high scattering efficiency. Especially at near infrared wavelengths (the $\lambda=1650$ and $2200$\,nm measurements) the scattering is dominated by ironsulfide grains.

The best fit model we present combines for the first time observations
of the SWS infrared spectrum ($7-44\,\mu$m), the LWS infrared spectrum
($44-120\,\mu$m) and the degree of linear polarization at several
wavelengths in the optical to near infrared. This results in a better constrained dust
model. The spectral structure of the thermal emission in the SWS part
of the spectrum provides crucial information on the composition of the
dust. The LWS part of the spectrum combined with the degree of linear
polarization provides information on the size of the dust grains. We
also constructed fits excluding some of the observations from the
model. An attempt to fit only the SWS part of the spectrum resulted in
an underestimate of the fraction of large grains, which in turn
resulted in a higher fraction of crystalline silicates, more comparable
to that found in, for example, \citet{2003A&A...401..577B}. Fitting
both the SWS and the LWS part of the spectrum without the linear
polarization resulted in a dust composition only slightly different (i.e. within the given error bars) from that presented in Table~\ref{tab:Fit}. The size distribution is affected more significantly. Also, using observations of the linear polarization at less wavelength points changed the parameters of the best fit model. For example, the best fit model when the observation of the degree of linear polarization at $\lambda=1650$ and $2200$\,nm are removed from the model contains no amorphous pyroxene and $8.5$\% amorphous silica. This can be explained in part by the interchangeability of olivine and pyroxene grains \citep[see also][]{vanBoekelMin2005}. The size distribution of this best fit model is changed such that the upper size limit of the amorphous olivine component is only $30\,\mu$m but the slope of the size distribution is more weigthed towards the larger grains, $\alpha=3.0$. The other parameters are only affected mildly.

The total dust mass in the SWS beam derived from the best fit model is
$4.6\cdot 10^9\,$kg. This dust mass is comparable to that estimated by
\citet{2003A&A...401..577B} which is $4.2\cdot 10^9\,$kg. Note that the dust mass derived in this way is a lower limit on
the real mass of the solid state material in the coma of Hale-Bopp
since the mass most likely resides predominantly in the very large
grains. Using data obtained at submillimeter wavelengths, which
provides information on millimeter sized dust grains,
\citet{1999AJ....117.1056J} derive a total dust mass of $\sim$$2\cdot
10^{11}\,$kg within a beam size comparable to that of the ISO SWS.

In our model the temperature of all dust species is determined
self-consistently. However, from the ratio of the strengths of the
forsterite features it can be seen that the temperature of the
forsterite grains is higher than would be determined from thermal
equilibrium calculations using pure forsterite grains
\citep{2002ApJ...580..579H}. Pure forsterite grains are not very
efficient absorbers in the UV and the optical part of the spectrum
where a large part of the solar energy is emitted. They are very
efficient emitters in the infrared. Therefore, pure forsterite grains
will be cold compared to other dust species. The fact that they are
observed to be relatively warm is probably an effect of thermal contact
between the various dust species \citep{2003A&A...401..577B}. An
aggregated structure where all dust species are in thermal contact is
also consistent with pictures of interplanetary dust particles. To
calculate the optical properties of an aggregated structure in a
completely consistent way is very computationally intensive which makes
it extremely difficult to examine the entire parameter space of dust
abundances and grains sizes. Some work on this has been done by
\citet{2003ApJ...595..522M} considering fixed values of the material
abundances. To simulate thermal contact we polluted the forsterite and
enstatite grains with 3\% of small metallic iron inclusions. We
calculated an effective refractive index using the Maxwell-Garnet
effective medium theory \citep[see e.g.][]{BohrenHuffman}. When we
pollute the crystalline silicates in this way, the equilibrium
temperature is in agreement with the observations.

\subsection{Dust composition and size distribution}

When discussing the size distribution of the dust we have to
consider the grain sizes that our analysis is sensitive to. For grains
smaller than a few micron ($\lesssim 3\,\mu$m) we have a strong
spectroscopic diagnostic. At relatively short wavelengths (around $\sim$10\,$\mu$m) we are mainly sensitive to the composition of these small 
grains. If we go to longer wavelengths, we are also sensitive to
the composition of larger grains. Using the SWS and LWS range from
$\lambda=7-120\,\mu$m, we have a spectroscopic diagnostic for the
composition of grains with a volume equivalent radius up to $\sim$10-15\,$\mu$m. Although larger grains do show spectral structure
\citep{2004A&A...413L..35M}, their emission efficiency is too low to be 
detected.

When we compare our best fit parameters with those obtained by others
\citep{1999P&SS...47..773B,1999EM&P...78..271G,1999ApJ...517.1034W,2000ApJ...538..428H,2002ApJ...580..579H,2003A&A...401..577B}
there are a few differences. First of all the amount of crystalline
silicates is much smaller than that found in most of these studies.
This is caused by the fact that previous studies mainly considered the
small (submicron sized) grain component. In our model, we find that in
order to reproduce the spectral features, the crystalline silicate
grains have to be very small and thus they have a high emission
efficiency. From, i.e. the LWS spectrum, we find that the amorphous
grains are relatively large, and thus emit less efficiently. Thus, if
one only considers the small grain component, the crystallinity is
increased with respect to our findings. The crystalline grains have to
be small in order to reproduce the ratios of the strengths of the
different emission features. However, these ratios are also influenced
by the temperature of the dust grains. Fortunately, we can distinguish
temperature effects from grain size effects in the following way. In
determining the absorption cross sections of dust particles the most
important parameter is $|mx|$ where $m$ is the complex refractive index
and $x=2\pi r/\lambda$ is the size parameter of the dust grain. When
$|mx|<<1$ the grains are in the Rayleigh domain and strong spectral
emission resonances occur. When $|mx|$ increases, the spectral features
decrease in strength. For very large grains ($|mx|>>1$) the emission
features will change into emission dips
\citep[see][]{2004A&A...413L..35M}. When the grain size is increased,
first the features caused by resonances with high values of
$|m/\lambda|$ become weaker and then the features with smaller values
of $|m/\lambda|$. This means that this effect depends on the wavelength
but also on the refractive index; the strongest features (with the
highest values of $|m|$) will go down first. However, when going from
high temperatures to low temperatures, the effect on the strength of
the features shifts with the maximum of the underlying Planck function,
so this effect only depends on the wavelength position of the features.
This difference allows us to distinguish between size and temperature
effects by carefully looking at the feature strength ratios.

\begin{figure}[!t]
\resizebox{\hsize}{!}{\includegraphics{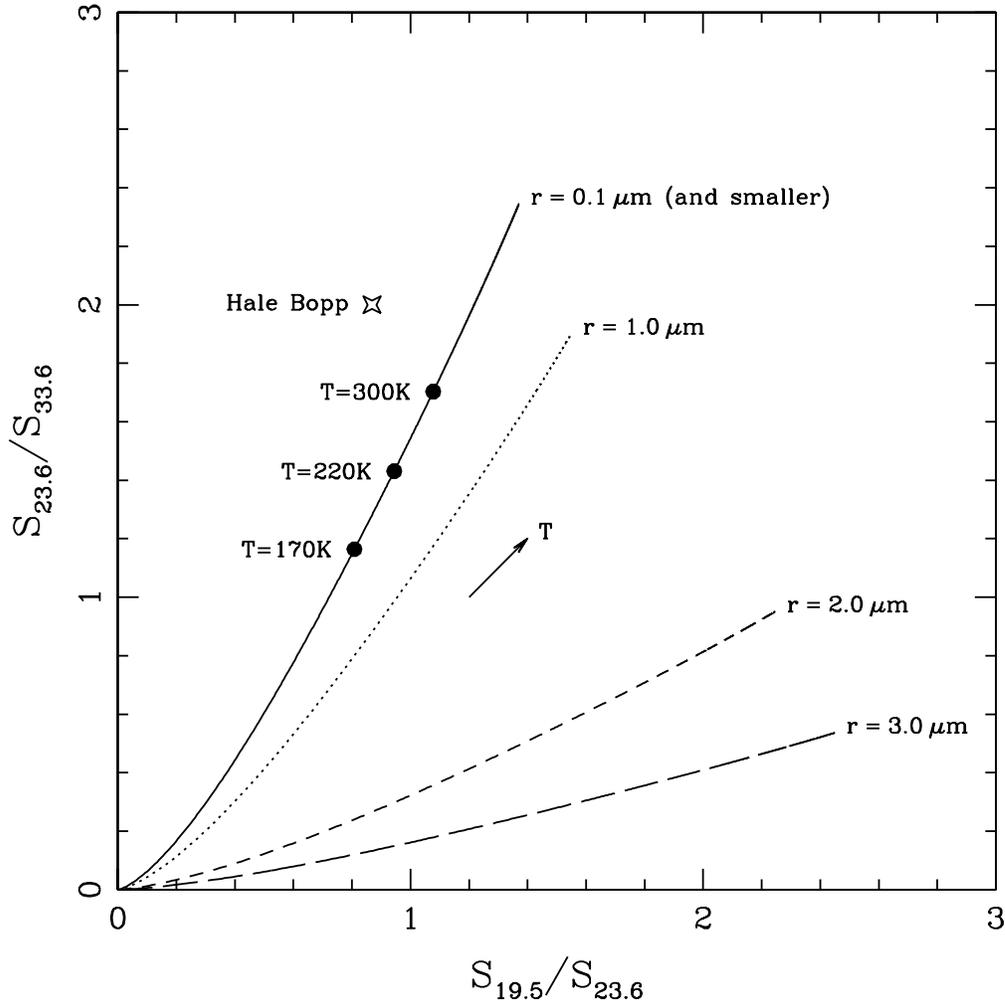}} 
\caption{The ratios
of the peak strengths at $19.5, 23.6$ and $33.6\,\mu$m of forsterite
calculated for various grain sizes and temperatures. The star indicates
the measured peak strength ratios of comet Hale-Bopp. Grains smaller
than $0.1\,\mu$m are in the Rayleigh limit. In this limit, the shape of
the spectrum is independent of the particle size.} 
\label{fig:peak ratios}
\end{figure}

To test if the ratios of the forsterite resonances can be explained
using temperature effects instead of a difference in grain size between
amorphous and crystalline silicates, we calculated the strengths of the
$19.5, 23.6$ and $33.6\,\mu$m features above the local continuum using
the DHS for different grain sizes and temperatures. These strengths are
denoted by $S_{19.5}, S_{23.6}$ and $S_{33.6}$ respectively. The
strength ratios are plotted in Fig.~\ref{fig:peak ratios} for various
grain sizes and temperatures. The temperatures were varied from
$10-1500$K. Also plotted in Fig.~\ref{fig:peak ratios} is the peak
ratio measured in the spectrum of Hale-Bopp. We see that there is no
corresponding set of temperature and grain size in Fig.~\ref{fig:peak
ratios} that reproduces the ratios measured in Hale-Bopp. To get the
best fit we need to go to very small and relatively warm forsterite
grains. The fact that the forsterite has to be warm also indicates that
the grains must be small and most probably are in thermal contact with
(at least) a strong absorbing material like carbon or iron.

The fact that we cannot find a set of temperature and grain size with
peak ratios in Fig.~\ref{fig:peak ratios} corresponding to those
observed in the infrared spectrum could be caused by the effect that in
the spectrum of Hale-Bopp the $33.6\,\mu$m feature is partly blended
with an enstatite feature resulting in a weaker feature. If we would
take this effect into account, the point would shift downward in
Fig.~\ref{fig:peak ratios}. Another explanation could be that the
forsterite in Hale-Bopp is slightly contaminated with iron, which
results in a slightly weaker $33.6\,\mu$m feature
\citep{1993MNRAS.264..654K}.

\begin{figure*}[!t]
\resizebox{\hsize}{!}{\includegraphics{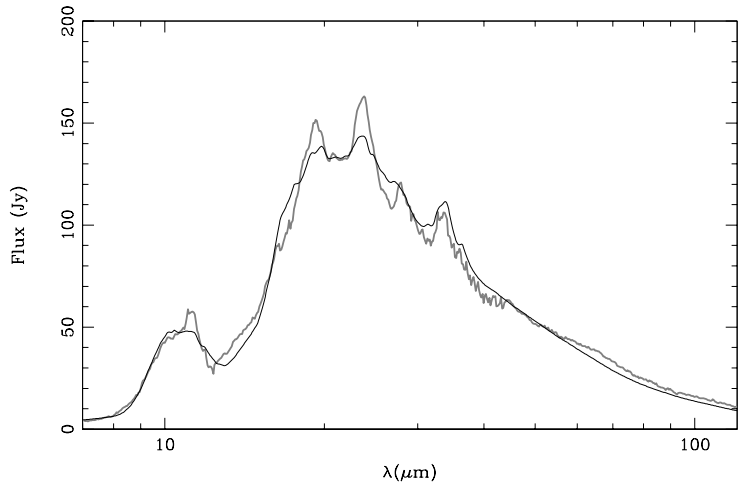}}
\caption{Infrared spectral energy distributions of the best possible fit when fixing the maximum volume equivalent radius, $r_\mathrm{max,cryst}$, of the crystalline component to $5\,\mu$m (black line). This fit to the SWS and LWS observations (gray line) therefore does not represent our best fit model (given in Fig.~\ref{fig:Spectral Fit}, which uses $r_\mathrm{max,cryst}\sim0.1\,\mu$m), but is of a much poorer quality. It is clear that large crystalline silicate grains cannot be present as they cannot reproduce the observed peak ratios of the resonances.} 
\label{fig:Spectral Fit Fix 5}
\end{figure*}

As a test we made a fit to the SWS and LWS spectra fixing the
upper size of the crystalline silicate grains. We have tried to make a
fit to the spectrum using $r_\mathrm{max, cryst}=2,5$ and $10\,\mu$m,
respectively, and varying all other parameters. The resulting model
spectra did not satisfactorily reproduce the measured ISO spectra. In
order to obtain a reasonable fit to the observed forsterite features,
we had to employ a power law for the size distribution more biased
towards small grains, $\beta=-3.6$. However, in all these model fits
the strength ratios of the forsterite features were poorly reproduced.
In Fig.~\ref{fig:Spectral Fit Fix 5} we plot the resulting best fit
model for the case when $r_\mathrm{max, cryst}=5\,\mu$m. We note that
the emission spectrum is not sensitive to the very large forsterite
grains. Although these grains still display significant spectral
structure \citep{2004A&A...413L..35M}, their emission efficiency is
small. Therefore, it is possible to have a bimodal size distribution of
crystalline silicates, in which only very small and very large crystals
are present while the intermediate sized grains are absent, and still
reproduce the infrared spectrum of Hale-Bopp. However, such a size
distribution is very unlikely.

When comparing our results to those obtained by other studies, we have
to be careful with the definition of crystallinity. Here, we mean by
crystallinity the mass fraction of crystalline silicates compared to
the total dust mass. In previous studies, sometimes the crystallinity
is defined as the mass fraction of crystalline silicates compared to
the total mass in the silicate component. In our fit, approximately
$40$\% of the dust mass is contained in non silicate materials like
carbon and ironsulfide. When comparing with other studies we have to
correct for this.

Since previous studies of the mineralogy of the dust in the coma of
Hale-Bopp focused on the SWS part of the spectrum, little information
was available on the large grain component. As can be seen from
Fig.~\ref{fig:Fit} the LWS spectrum (longward of $\sim$44\,$\mu$m)
displays a long wavelength slope that can only be explained using large
grains at approximately blackbody temperature \citep[see
also][]{1998A&A...339L...9L}. We have just shown that this large grain
component cannot be crystalline. Since the large grains contain most of
the mass, the resulting abundance of crystalline silicates is small,
$4.3$\% (Table~\ref{tab:Fit}). We should note here that part of the
crystalline silicates are probably not spectroscopically detectable
since these are hidden inside large amorphous silicate or carbon
grains. Therefore the $4.3$\% crystallinity we derive is a lower limit.
We can make a quantitative estimate of the true fraction of crystalline
grains by only considering grains smaller than a certain size, for
which we assume that we observe the properties of the entire grain. In
order to assure that the resulting mixture still has solar abundances,
we fix the total abundance of silicates (crystalline and amorphous
olivine and pyroxene). Considering only the silicate component, we
compute the crystalline over amorphous ratio excluding the larger
silicate grains. We then use this crystalline over amorphous ratio to
compute the fraction of crystalline silicates in the total mixture. We
choose a volume equivalent radius of $10\,\mu$m as the maximum grain
size, which represents approximately half of the total dust mass. The
crystallinity derived in this way depends on the assumption of this
grain size, and detailed calculations are required to obtain a better
estimate of the appropriate size. We thus derive a crystallinity of
$\sim$7.5\% ($\sim$5\% of crystalline forsterite and $\sim$2.5\% of
crystalline enstatite). Note that most previous studies on the SWS
spectrum of Hale-Bopp only included forsterite, no enstatite. In
contrast with our findings, these studies typically find forsterite
fractions of $\sim$20-30\%. The fraction of crystalline silicates in
our best fit model increases when we consider only smaller grains. The
submicron grain component contains approximately $30$\% forsterite and
$14$\% enstatite. This fraction of forsterite is consistent with the
results from previous studies that focused on the submicron grain
component \citep[see e.g.][]{1999P&SS...47..773B}.

It can be noted from Table~\ref{tab:Fit} that the dust components with
a high abundance reside in relatively large grains. Although the large
grains contribute much to the total dust mass, they contribute only
little to the infrared emission at short wavelengths. Therefore, we
conclude that the long wavelength part of the spectrum and the degree
of linear polarization contain crucial information on the large dust
grains, and, therefore, on a large fraction of the total dust mass.

We can calculate the effective radius and variance, $r_\mathrm{eff}$
and $v_\mathrm{eff}$, of the dust grains in our model using
Eqs.~(\ref{eq:reff}) and (\ref{eq:veff}). For our best fit model
$r_\mathrm{eff}=1.0\,\mu$m and $v_\mathrm{eff}=16.8$. Typical values for
the effective volume equivalent radii computed from the size
distributions found in other studies are $0.5\,\mu$m
\citep{2003A&A...401..577B} and $0.6\,\mu$m
\citep{2003ApJ...595..522M}. The larger $r_\mathrm{eff}$ that we find
is most likely due to the additional information contained in the long
wavelength slope. Employing a simple model for porous dust grains,
\citet{1998ApJ...498L..83L} argue that the average grain size in
Hale-Bopp might be much larger ($\sim$8-25\,$\mu$m) and that the
fraction of submicron grains might be negligible. Perhaps cometary dust
grains are indeed very fluffy aggregates of small particles. However,
it is probably difficult to discriminate between a large fluffy grain
and a large number of small separate grains using spectroscopy or
observations of the degree of linear polarization.

The fact that the most abundant dust species reside in the largest
grains is a natural consequence of the formation mechanism of the
larger dust grains. Larger dust grains are believed to form by
coagulation of smaller grains. The materials that are more abundant,
can form large clusters of this material, whereas the less abundant
species may be distributed in the aggregate as separate monomers. If
the aggregate is very fluffy, the low abundance of the latter species
causes the distance between the inclusions to be on average relatively
large. Thus, one could argue that these inclusions will not interact
very strongly with each other, and their resulting optical properties
will be those of small particles. On the other hand, the larger
clusters formed by the more abundant dust species will produce optical
properties more like those of large grains. This hypothesis may be
tested by simulations of the optical properties of fluffy composite
particles with extreme abundance differences.

We conclude that the crystalline silicates in the coma of comet
Hale-Bopp are submicron sized. This is much smaller than the typical
grain size found for the other dust components. In agreement with these
findings, the crystalline silicates found in IDPs are predominantly
submicron sized \citep{Bradley}. This difference in grain size between
the amorphous and crystalline components results in a relatively low
overall abundance of crystalline silicates. The abundance of crystalline
silicates and the difference in grain size are important constraints
when considering models of the processing and dynamical history of dust
in the early solar system.

\section{Discussion: Origin and evolution of cometary dust}
\label{sec:discussion}

Since comets are small bodies that have been frozen all of their
lifetime, the comet material is expected to have undergone little
processing since the comet's formation. In larger bodies, like planets,
the material has experienced severe parent body processing erasing the
information on the original dust grains. Therefore, comets contain a
unique diagnostic of the mineralogy of the dust in the protoplanetary
disk at times when the planets and comets were formed. The fraction of
crystalline silicates in the diffuse interstellar medium (ISM) is very
low. \citet{2001ApJ...550L.213L} derive an upper limit for the
crystallinity of the dust in the ISM of $5$\%, while
\citet{2004ApJ...609..826K} even derive an upper limit of $0.4$\%. In
addition, the crystallinity of the ISM dust derived by \citet{vanBoekelMin2005} is $\sim$1\%. This implies
that the dust in Hale-Bopp has undergone at least some processing in
the solar nebula, before it got incorporated into the comet. In order
to form crystalline silicates, the amorphous silicates have to be
heated above the glass temperature of $\sim$1000\,K. Comets form in
the outer regions of the disk where the temperatures are low enough for
water ice to exist $\lesssim 160$\,K. There are two possible ways to
explain the presence of crystalline silicates in cometary dust grains.
The first is that the amorphous silicates are crystallized by thermal
annealing in the hot inner regions of the protoplanetary disk. The
crystalline silicates are then mixed out to the regions where the
comets form. Another possibility is that the crystalline silicates are
produced locally. Processes that have been proposed for this are, for
example, shock annealing \citep{2002ApJ...565L.109H} and lightning
\citep{1998A&A...331..121P, 2000Icar..143...87D}. Recent evidence for
radial mixing in protoplanetary disks is presented in \citet{2004Natur.432..479V} on the basis of
interferometric measurements of the dust in the inner disk regions. In
that paper it is found that the crystallinity in the inner disk regions
is higher than that in the outer disk regions. This difference, along
with a varying forsterite over enstatite ratio, is consistent with
predictions from radial mixing models \citep{2004A&A...413..571G}.

Several studies have tried to explain the presence of crystalline
silicates in comets by considering radial mixing. In these studies only
the evolution of the silicate component is computed. For comparison
with our computations, we therefore have to correct for the additional
components included in the fit. The fraction of crystalline silicates
in the silicate component (amorphous and crystalline olivine and
pyroxene) is approximately $12.5$\%. In \citet{2004A&A...413..571G} a
detailed model for the protosolar disk is presented in which dust
chemistry, thermal annealing and radial mixing are incorporated. The
computations in this paper are for a stationary model, so an
equilibrium situation is calculated. The crystallinity that follows
from this detailed model depends strongly on the distance to the star.
Close to the star, the crystallinity is very high ($\sim$70\% at
$3$\,AU), while in the comet forming region ($\sim$20\,AU) the
crystalline fraction equals approximately $24$\% ($8$\% crystalline
olivine and $16$\% crystalline pyroxene). The author concludes that the
results from the stationary model cannot be extrapolated to distances
beyond $20$\,AU, and can only be considered an approximation for the
mineral composition of the inner $20$\,AU at several times $10^5$
years. \citet{2002A&A...385..181W} present a time depend model of
radial mixing. From the results of this paper, it is apparent that the
equilibrium situation as computed in \citet{2004A&A...413..571G} is
already attained after $\sim$$10^5$ years. This is slightly shorter than
the anticipated typical timescale for comet formation \citep[which is a
few times $10^5$ years, see][]{1997Icar..127..290W}. The crystallinity
computed by \citet{2002A&A...385..181W} at a distance of $30$\,AU is
approximately $7$\% after $10^6$ years. \citet{2002A&A...384.1107B}
also present a model describing time-dependent radial mixing in a
protoplanetary disk. In this paper, three different solar nebula models
are presented, a warm, a nominal, and a cold model referring to the
temperature structure in the solar nebula. This temperature structure
is set by the viscosity parameter $\alpha$; the higher the values of
$\alpha$ the lower the temperature. They derive an extremely well mixed
nebula, in which, after $\sim$$10^6$ years, the crystallinity at
distances $>10$\,AU is independent of the distance. They arrive at a
final crystallinity in the outer solar system ($>10$\,AU) of
approximately $58$, $12$ and $2$\% according to the warm, nominal, and
cold solar nebula model. Note that the parameters chosen by
\citet{2002A&A...385..181W} and \citet{2004A&A...413..571G} correspond
to the warm to nominal solar nebula model of
\citet{2002A&A...384.1107B}. The predicted crystallinity in the comet
forming region from both \citet{2004A&A...413..571G} and
\citet{2002A&A...384.1107B} shows that thermal annealing and radial
mixing are more than sufficiently efficient mechanisms to explain the
crystalline silicates in Hale-Bopp. The most important free parameter
in the above models is the viscosity parameter. The crystallinity of
cometary dust can be used to constrain this parameter, providing a
better insight in the dynamics of protoplanetary disks. While the
crystallinity derived for Hale-Bopp in previous studies could only be
explained employing a viscosity parameter that is representative for a
warm solar nebula model, the crystallinity we derive for Hale-Bopp is
consistent with a viscosity parameter that is typical for an
approximately nominal solar nebula and a formation distance of some
$30$\,AU from the central star.

The crystalline silicates we find are all submicron sized, consistent
with studies of IDPs, in which the crystalline inclusions are
predominantly submicron sized \citep{Bradley}. There are two possible
explanations for this. The first explanation is that the crystalline
silicates are formed before efficient grain growth sets in. Due to the
low crystallinity this would result in a dust grain that has the
crystalline silicates scattered in the aggregate as small separate
inclusions. However, this possibility can likely be excluded on the
basis of the results presented by \citet{vanBoekelMin2005}. 
From the analysis of a large sample of
protoplanetary disks, it is concluded that grain growth occurs before
efficient crystallization sets in. Another explanation might be that
the mechanism that produces the crystalline silicates in the comet
forming region is more efficient for small grains than for large
grains. Both local flash heating events (like shock annealing and
lightning) as well as radial mixing are more efficient for small
grains. For the local production mechanisms this is due to the fact
that small grains are more easily heated than large grains. In the
radial mixing models this is due to the fact that small grains more
easily couple to the gas, and are thus more easily mixed outwards by,
for example, turbulent radial mixing. It is, however, unclear if the
size dependencies of the various models are strong enough to explain
the absence of large crystalline silicate grains in the coma of
Hale-Bopp.

\section{Conclusions}
\label{sec:conclusion}

We have successfully modeled the thermal emission and the degree of
linear polarization of radiation scattered by grains in the coma of
comet Hale-Bopp. Our method has the following important characteristics
relative to previous studies.
\begin{itemize}
\item Both grain size and grain shape effects are taken into account in the
calculations of the optical properties.
\item The parameters of the best fit model are determined using an objective
least squares fitting routine.
\item The abundances of the chemical elements observed in interplanetary dust particles and meteorites could be used as constraints for the model.
\item The resulting model is consistent with the infrared emission spectrum
observed in the wavelength range $7-120\,\mu$m and with observations of the
degree of linear polarization at various phase angles and twelve different
wavelengths in the optical to near infrared part of the spectrum.
\end{itemize}
To model the effects of grain shape on the optical properties, we
employed the distribution of hollow spheres. In this distribution we
average over the volume fraction occupied by the central vacuum
inclusion while preserving the material volume of the particles. We
showed that this shape distribution is successful in reproducing the
observed properties of cometary grains.

We deduced from the ratios of the strengths of various forsterite
features in the observed spectrum of Hale-Bopp that the crystalline
silicate grains have a volume equivalent radius $r\lesssim 1\,\mu$m.
This is much smaller than the typical grain size of the other dust
components and is in agreement with the sizes of the crystalline
silicate inclusions found in fluffy interplanetary dust particles. The
crystalline inclusions in these grains are predominantly submicron
sized \citep{Bradley}.

The long wavelength observations showed that most of the mass resides in relatively large grains. The lack of large crystalline silicate grains in our model thus implies that the amount of mass in this component is small.
Our best fit model has a relative amount of crystalline silicates that
is significantly lower than found in previous studies of the infrared
spectrum. If we consider only the grains with a volume equivalent
radius smaller than $10\,\mu$m, the fraction of the total dust mass
contained in crystalline silicates is only $\sim$7.5\%. The fraction
of crystalline silicates in the silicate component (both amorphous and
crystalline olivine and pyroxene) is $\sim$12.5\%. This crystallinity
can easily be produced by models in which the crystalline silicates are
formed close to the Sun by thermal annealing and then mixed outwards to
the comet forming region ($\sim$20-30\,AU). The crystallinity derived
by us for comet Hale-Bopp is in agreement with these models assuming an
approximately nominal model of the protosolar nebula and a formation of
the comet at a distance of $\sim$30\,AU from the Sun. This
crystallinity is also in agreement with that found in interplanetary
dust particles.

\section*{Acknowledgments}
It is a pleasure to express our gratitude to M.~S.~Hanner, C.~Dijkstra and L.~Kolokolova for enlightening discussions. We are
grateful to J.~Crovisier for providing us with the reduced data of
the LWS spectrum of Hale-Bopp.


\begin{thebibliography}{64}
\expandafter\ifx\csname natexlab\endcsname\relax\def\natexlab#1{#1}\fi
\expandafter\ifx\csname url\endcsname\relax
  \def\url#1{\texttt{#1}}\fi
\expandafter\ifx\csname urlprefix\endcsname\relax\def\urlprefix{URL }\fi

\bibitem[{{Aden} and {Kerker}(1951)}]{Aden1951}
{Aden}, A.~L., {Kerker}, M., 1951. {Scattering of electromagnetic waves from
  two concentric spheres}. Journal of Applied Physics 22, 1242--1246.

\bibitem[{{Begemann} et~al.(1994){Begemann}, {Dorschner}, {Henning},
  {Mutschke}, and {Thamm}}]{1994ApJ...423L..71B}
{Begemann}, B., {Dorschner}, J., {Henning}, T., {Mutschke}, H., {Thamm}, E.,
  Mar. 1994. {A laboratory approach to the interstellar sulfide dust problem}.
  \apjl 423, L71--L74.

\bibitem[{{Bockel{\' e}e-Morvan} et~al.(2002){Bockel{\' e}e-Morvan}, {Gautier},
  {Hersant}, {Hur{\' e}}, and {Robert}}]{2002A&A...384.1107B}
{Bockel{\' e}e-Morvan}, D., {Gautier}, D., {Hersant}, F., {Hur{\' e}}, J.-M.,
  {Robert}, F., Mar. 2002. {Turbulent radial mixing in the solar nebula as the
  source of crystalline silicates in comets.} \aap 384, 1107--1118.

\bibitem[{{Bockel{\' e}e-Morvan} et~al.(2000){Bockel{\' e}e-Morvan}, {Lis},
  {Wink}, {Despois}, {Crovisier}, {Bachiller}, {Benford}, {Biver}, {Colom},
  {Davies}, {G{\' e}rard}, {Germain}, {Houde}, {Mehringer}, {Moreno},
  {Paubert}, {Phillips}, and {Rauer}}]{2000A&A...353.1101B}
{Bockel{\' e}e-Morvan}, D., {Lis}, D.~C., {Wink}, J.~E., {Despois}, D.,
  {Crovisier}, J., {Bachiller}, R., {Benford}, D.~J., {Biver}, N., {Colom}, P.,
  {Davies}, J.~K., {G{\' e}rard}, E., {Germain}, B., {Houde}, M., {Mehringer},
  D., {Moreno}, R., {Paubert}, G., {Phillips}, T.~G., {Rauer}, H., Jan. 2000.
  {New molecules found in comet C/1995 O1 (Hale-Bopp). Investigating the link
  between cometary and interstellar material}. \aap 353, 1101--1114.

\bibitem[{{Bohren} and {Huffman}(1983)}]{BohrenHuffman}
{Bohren}, C.~F., {Huffman}, D.~R., 1983. {Absorption and scattering of light by
  small particles}. New York: Wiley.

\bibitem[{{Bouwman} et~al.(2003){Bouwman}, {de Koter}, {Dominik}, and
  {Waters}}]{2003A&A...401..577B}
{Bouwman}, J., {de Koter}, A., {Dominik}, C., {Waters}, L.~B.~F.~M., Apr. 2003.
  {The origin of crystalline silicates in the Herbig Be star HD 100546 and in
  comet Hale-Bopp}. \aap 401, 577--592.

\bibitem[{{Bouwman} et~al.(2001){Bouwman}, {Meeus}, {de Koter}, {Hony},
  {Dominik}, and {Waters}}]{2001A&A...375..950B}
{Bouwman}, J., {Meeus}, G., {de Koter}, A., {Hony}, S., {Dominik}, C.,
  {Waters}, L.~B.~F.~M., Sep. 2001. {Processing of silicate dust grains in
  Herbig Ae/Be systems}. \aap 375, 950--962.

\bibitem[{{Bradley} et~al.(1983){Bradley}, {Brownlee}, and
  {Veblen}}]{1983Natur.301..473B}
{Bradley}, J.~P., {Brownlee}, D.~E., {Veblen}, D.~R., Feb. 1983. {Pyroxene
  whiskers and platelets in interplanetary dust - Evidence of vapour phase
  growth}. \nat 301, 473--477.

\bibitem[{Bradley et~al.(1999)Bradley, Snow, Brownlee, and Hanner}]{Bradley}
Bradley, J.~P., Snow, T.~P., Brownlee, D.~E., Hanner, M.~S., 1999. Sollid
  Interstellar Matter: The ISO Revolution. No.~11 in Centre de Physique des
  Houches. Springer, Berlin, p. 297.

\bibitem[{{Brucato} et~al.(1999){Brucato}, {Colangeli}, {Mennella}, {Palumbo},
  and {Bussoletti}}]{1999P&SS...47..773B}
{Brucato}, J.~R., {Colangeli}, L., {Mennella}, V., {Palumbo}, P., {Bussoletti},
  E., Jun. 1999. {Silicates in Hale-Bopp: hints from laboratory studies}.
  \planss 47, 773--779.

\bibitem[{{Charbonneau}(1995)}]{1995ApJS..101..309C}
{Charbonneau}, P., Dec. 1995. {Genetic Algorithms in Astronomy and
  Astrophysics}. \apjs 101, 309--+.

\bibitem[{{Crovisier} et~al.(1997){Crovisier}, {Leech}, {Bockelee-Morvan},
  {Brooke}, {Hanner}, {Altieri}, {Keller}, and
  {Lellouch}}]{1997Sci...275.1904C}
{Crovisier}, J., {Leech}, K., {Bockelee-Morvan}, D., {Brooke}, T.~Y., {Hanner},
  M.~S., {Altieri}, B., {Keller}, H.~U., {Lellouch}, E., 1997. {The spectrum of
  Comet Hale-Bopp (C/1995 01) observed with the Infrared Space Observatory at
  2.9 AU from the Sun}. Science 275, 1904--1907.

\bibitem[{{Desch} and {Cuzzi}(2000)}]{2000Icar..143...87D}
{Desch}, S.~J., {Cuzzi}, J.~N., Jan. 2000. {The Generation of Lightning in the
  Solar Nebula}. Icarus 143, 87--105.

\bibitem[{{Dorschner} et~al.(1995){Dorschner}, {Begemann}, {Henning}, {J{\"
  a}ger}, and {Mutschke}}]{1995A&A...300..503D}
{Dorschner}, J., {Begemann}, B., {Henning}, T., {J{\" a}ger}, C., {Mutschke},
  H., Aug. 1995. {Steps toward interstellar silicate mineralogy. II. Study of
  Mg-Fe-silicate glasses of variable composition.} \aap 300, 503--+.

\bibitem[{{Fabian} et~al.(2000){Fabian}, {J{\" a}ger}, {Henning}, {Dorschner},
  and {Mutschke}}]{2000A&A...364..282F}
{Fabian}, D., {J{\" a}ger}, C., {Henning}, T., {Dorschner}, J., {Mutschke}, H.,
  Dec. 2000. {Steps toward interstellar silicate mineralogy. V. Thermal
  Evolution of Amorphous Magnesium Silicates and Silica}. \aap 364, 282--292.

\bibitem[{{Forrest} et~al.(2004){Forrest}, {Sargent}, {Furlan}, {D'Alessio},
  {Calvet}, {Hartmann}, {Uchida}, {Green}, {Watson}, {Chen}, {Kemper},
  {Keller}, {Sloan}, {Herter}, {Brandl}, {Houck}, {Barry}, {Hall}, {Morris},
  {Najita}, and {Myers}}]{2004ApJS..154..443F}
{Forrest}, W.~J., {Sargent}, B., {Furlan}, E., {D'Alessio}, P., {Calvet}, N.,
  {Hartmann}, L., {Uchida}, K.~I., {Green}, J.~D., {Watson}, D.~M., {Chen},
  C.~H., {Kemper}, F., {Keller}, L.~D., {Sloan}, G.~C., {Herter}, T.~L.,
  {Brandl}, B.~R., {Houck}, J.~R., {Barry}, D.~J., {Hall}, P., {Morris}, P.~W.,
  {Najita}, J., {Myers}, P.~C., Sep. 2004. {Mid-infrared Spectroscopy of Disks
  around Classical T Tauri Stars}. \apjs 154, 443--447.

\bibitem[{{Gail}(2004)}]{2004A&A...413..571G}
{Gail}, H.-P., Jan. 2004. {Radial mixing in protoplanetary accretion disks. IV.
  Metamorphosis of the silicate dust complex}. \aap 413, 571--591.

\bibitem[{{Galdemard} et~al.(1999){Galdemard}, {Lagage}, {Dubreuil}, {Jouan},
  {Masse}, {Pantin}, and {Bockel{\' e}e-Morvan}}]{1999EM&P...78..271G}
{Galdemard}, P., {Lagage}, P.~O., {Dubreuil}, D., {Jouan}, R., {Masse}, P.,
  {Pantin}, E., {Bockel{\' e}e-Morvan}, D., 1999. {Mid-Infrared Spectro-Imaging
  Observations Of Comet Hale-Bopp}. Earth Moon and Planets 78, 271--277.

\bibitem[{{Ganesh} et~al.(1998){Ganesh}, {Joshi}, {Baliyan}, and
  {Deshpande}}]{1998A&AS..129..489G}
{Ganesh}, S., {Joshi}, U.~C., {Baliyan}, K.~S., {Deshpande}, M.~R., May 1998.
  {Polarimetric observations of the comet Hale-Bopp}. \aaps 129, 489--493.

\bibitem[{{Geiss}(1987)}]{1987A&A...187..859G}
{Geiss}, J., Nov. 1987. {Composition measurements and the history of cometary
  matter}. \aap 187, 859--866.

\bibitem[{{Grevesse} and {Sauval}(1998)}]{1998SSRv...85..161G}
{Grevesse}, N., {Sauval}, A.~J., Aug. 1998. {Standard Solar Composition}. Space
  Science Reviews 85, 161--174.

\bibitem[{{Hanner} et~al.(1999){Hanner}, {Gehrz}, {Harker}, {Hayward}, {Lynch},
  {Mason}, {Russell}, {Williams}, {Wooden}, and
  {Woodward}}]{1999EM&P...79..247H}
{Hanner}, M.~S., {Gehrz}, R.~D., {Harker}, D.~E., {Hayward}, T.~L., {Lynch},
  D.~K., {Mason}, C.~C., {Russell}, R.~W., {Williams}, D.~M., {Wooden}, D.~H.,
  {Woodward}, C.~E., 1999. {Thermal Emission from the Dust Coma of Comet
  Hale-Bopp and the Composition of the Silicate Grains}. Earth Moon and Planets
  79, 247--264.

\bibitem[{{Harker} and {Desch}(2002)}]{2002ApJ...565L.109H}
{Harker}, D.~E., {Desch}, S.~J., Feb. 2002. {Annealing of Silicate Dust by
  Nebular Shocks at 10 AU}. \apjl 565, L109--L112.

\bibitem[{{Harker} et~al.(2002){Harker}, {Wooden}, {Woodward}, and
  {Lisse}}]{2002ApJ...580..579H}
{Harker}, D.~E., {Wooden}, D.~H., {Woodward}, C.~E., {Lisse}, C.~M., Nov. 2002.
  {Grain Properties of Comet C/1995 O1 (Hale-Bopp)}. \apj 580, 579--597.

\bibitem[{{Hasegawa} et~al.(1999){Hasegawa}, {Ichikawa}, {Abe}, {Hamamura},
  {Ohnishi}, and {Watanabe}}]{1999EM&P...78..353H}
{Hasegawa}, H., {Ichikawa}, T., {Abe}, S., {Hamamura}, S., {Ohnishi}, K.,
  {Watanabe}, J., 1999. {Near-Infrared Photometric and Polarimetric
  Observations Of Comet Hale-Bopp}. Earth Moon and Planets 78, 353--358.

\bibitem[{{Hayward} et~al.(2000){Hayward}, {Hanner}, and
  {Sekanina}}]{2000ApJ...538..428H}
{Hayward}, T.~L., {Hanner}, M.~S., {Sekanina}, Z., Jul. 2000. {Thermal Infrared
  Imaging and Spectroscopy of Comet Hale-Bopp (C/1995 O1)}. \apj 538, 428--455.

\bibitem[{{Henning} et~al.(1996){Henning}, {Chan}, and
  {Assendorp}}]{1996A&A...312..511H}
{Henning}, T., {Chan}, S.~J., {Assendorp}, R., Aug. 1996. {The nature of
  objects with a 21-{$\mu$}m feature.} \aap 312, 511--520.

\bibitem[{{Irvine} et~al.(2000){Irvine}, {Senay}, {Lovell}, {Matthews},
  {McGonagle}, and {Meier}}]{2000Icar..143..412I}
{Irvine}, W.~M., {Senay}, M., {Lovell}, A.~J., {Matthews}, H.~E., {McGonagle},
  D., {Meier}, R., Feb. 2000. {NOTE: Detection of Nitrogen Sulfide in Comet
  Hale-Bopp}. Icarus 143, 412--414.

\bibitem[{{J{\" a}ger} et~al.(1998){J{\" a}ger}, {Molster}, {Dorschner},
  {Henning}, {Mutschke}, and {Waters}}]{1998A&A...339..904J}
{J{\" a}ger}, C., {Molster}, F.~J., {Dorschner}, J., {Henning}, T., {Mutschke},
  H., {Waters}, L.~B.~F.~M., Nov. 1998. {Steps toward interstellar silicate
  mineralogy. IV. The crystalline revolution}. \aap 339, 904--916.

\bibitem[{{Jewitt} and {Matthews}(1999)}]{1999AJ....117.1056J}
{Jewitt}, D., {Matthews}, H., Feb. 1999. {Particulate Mass Loss from Comet
  Hale-Bopp}. \aj 117, 1056--1062.

\bibitem[{{Jockers} et~al.(1999){Jockers}, {Rosenbush}, {Bonev}, and
  {Credner}}]{1999EM&P...78..373J}
{Jockers}, K., {Rosenbush}, V.~K., {Bonev}, T., {Credner}, T., 1999. {Images of
  Polarization and Colour in the Inner Coma of Comet Hale-Bopp}. Earth Moon and
  Planets 78, 373--379.

\bibitem[{{Jones} and {Gehrz}(2000)}]{2000Icar..143..338J}
{Jones}, T.~J., {Gehrz}, R.~D., Feb. 2000. {Infrared Imaging Polarimetry of
  Comet C/1995 01 (Hale-Bopp)}. Icarus 143, 338--346.

\bibitem[{{Kahnert} et~al.(2002){Kahnert}, {Stamnes}, and
  {Stamnes}}]{2002JQSRT..74..167K}
{Kahnert}, F.~M., {Stamnes}, J.~J., {Stamnes}, K., Jul. 2002. {Using simple
  particle shapes to model the Stokes scattering matrix of ensembles of
  wavelength-sized particles with complex shapes: possibilities and
  limitations}. Journal of Quantitative Spectroscopy and Radiative Transfer 74,
  167--182.

\bibitem[{{Kahnert}(2004)}]{2004JQSRT..85..231K}
{Kahnert}, M., May 2004. {Reproducing the optical properties of fine desert
  dust aerosols using ensembles of simple model particles}. Journal of
  Quantitative Spectroscopy and Radiative Transfer 85, 231--249.

\bibitem[{{Kemper} et~al.(2004){Kemper}, {Vriend}, and
  {Tielens}}]{2004ApJ...609..826K}
{Kemper}, F., {Vriend}, W.~J., {Tielens}, A.~G.~G.~M., Jul. 2004. {The Absence
  of Crystalline Silicates in the Diffuse Interstellar Medium}. \apj 609,
  826--837.

\bibitem[{{Koike} et~al.(1993){Koike}, {Shibai}, and
  {Tuchiyama}}]{1993MNRAS.264..654K}
{Koike}, C., {Shibai}, H., {Tuchiyama}, A., Oct. 1993. {Extinction of Olivine
  and Pyroxene in the Mid Infrared and Far Infrared}. \mnras 264, 654--+.

\bibitem[{{Kolokolova} et~al.(2004){Kolokolova}, {Hanner}, {Levasseur-Regourd},
  and {Gustafson}}]{Kolokolova2004}
{Kolokolova}, L., {Hanner}, M.~S., {Levasseur-Regourd}, A.~C., {Gustafson}, B.
  {\.A}.~S., 2004. {Physical properties of cometary dust from light scattering
  and thermal emission}. Comets II. Arizona Press.

\bibitem[{{Lellouch} et~al.(1998){Lellouch}, {Crovisier}, {Lim},
  {Bockelee-Morvan}, {Leech}, {Hanner}, {Altieri}, {Schmitt}, {Trotta}, and
  {Keller}}]{1998A&A...339L...9L}
{Lellouch}, E., {Crovisier}, J., {Lim}, T., {Bockelee-Morvan}, D., {Leech}, K.,
  {Hanner}, M.~S., {Altieri}, B., {Schmitt}, B., {Trotta}, F., {Keller}, H.~U.,
  Nov. 1998. {Evidence for water ice and estimate of dust production rate in
  comet Hale-Bopp at 2.9 AU from the Sun}. \aap 339, L9--L12.

\bibitem[{{Li} and {Draine}(2001)}]{2001ApJ...550L.213L}
{Li}, A., {Draine}, B.~T., Apr. 2001. {On Ultrasmall Silicate Grains in the
  Diffuse Interstellar Medium}. \apjl 550, L213--L217.

\bibitem[{{Li} and {Greenberg}(1998)}]{1998ApJ...498L..83L}
{Li}, A., {Greenberg}, J.~M., May 1998. {From Interstellar Dust to Comets:
  Infrared Emission from Comet Hale-Bopp (C/1995 O1)}. \apjl 498, L83--+.

\bibitem[{{Li} et~al.(2002){Li}, {Greenberg}, and {Zhao}}]{2002MNRAS.334..840L}
{Li}, A., {Greenberg}, J.~M., {Zhao}, G., Aug. 2002. {Modelling the
  astronomical silicate features - I. On the spectrum subtraction method}.
  \mnras 334, 840--846.

\bibitem[{{Manset} and {Bastien}(2000)}]{2000Icar..145..203M}
{Manset}, N., {Bastien}, P., May 2000. {Polarimetric Observations of Comets
  C/1995 O1 Hale-Bopp and C/1996 B2 Hyakutake}. Icarus 145, 203--219.

\bibitem[{{Mason} et~al.(2001){Mason}, {Gehrz}, {Jones}, {Woodward}, {Hanner},
  and {Williams}}]{2001ApJ...549..635M}
{Mason}, C.~G., {Gehrz}, R.~D., {Jones}, T.~J., {Woodward}, C.~E., {Hanner},
  M.~S., {Williams}, D.~M., Mar. 2001. {Observations of Unusually Small Dust
  Grains in the Coma of Comet Hale-Bopp C/1995 O1}. \apj 549, 635--646.

\bibitem[{{Min} et~al.(2004){Min}, {Dominik}, and
  {Waters}}]{2004A&A...413L..35M}
{Min}, M., {Dominik}, C., {Waters}, L.~B.~F.~M., Jan. 2004. {Spectroscopic
  diagnostic for the mineralogy of large dust grains}. \aap 413, L35--L38.

\bibitem[{{Min} et~al.(2003){Min}, {Hovenier}, and {de
  Koter}}]{2003A&A...404...35M}
{Min}, M., {Hovenier}, J.~W., {de Koter}, A., Jun. 2003. {Shape effects in
  scattering and absorption by randomly oriented particles small compared to
  the wavelength}. \aap 404, 35--46.

\bibitem[{{Min} et~al.(2005){Min}, {Hovenier}, and {de Koter}}]{MinHollow}
{Min}, M., {Hovenier}, J.~W., {de Koter}, A., 2005. {Modeling optical
  properties of cosmic dust grains using a distribution of hollow spheres}.
  \aap 432, 909--920.

\bibitem[{Mishchenko et~al.(2002)Mishchenko, Travis, and
  Lacis}]{MishTravis2002}
Mishchenko, M.~I., Travis, L.~D., Lacis, A.~A., 2002. Scattering, Absorption
  and Emission of Light by Small Particles. Cambridge University Press,
  Cambridge.

\bibitem[{{Molster} et~al.(1999){Molster}, {Waters}, {Trams}, {Van Winckel},
  {Decin}, {van Loon}, {J{\" a}ger}, {Henning}, {K{\" a}ufl}, {de Koter}, and
  {Bouwman}}]{1999A&A...350..163M}
{Molster}, F.~J., {Waters}, L.~B.~F.~M., {Trams}, N.~R., {Van Winckel}, H.,
  {Decin}, L., {van Loon}, J.~T., {J{\" a}ger}, C., {Henning}, T., {K{\"
  a}ufl}, H.-U., {de Koter}, A., {Bouwman}, J., Oct. 1999. {The composition and
  nature of the dust shell surrounding the binary AFGL 4106}. \aap 350,
  163--180.

\bibitem[{{Moreno} et~al.(2003){Moreno}, {Mu{\~ n}oz}, {Vilaplana}, and
  {Molina}}]{2003ApJ...595..522M}
{Moreno}, F., {Mu{\~ n}oz}, O., {Vilaplana}, R., {Molina}, A., Sep. 2003.
  {Irregular Particles in Comet C/1995 O1 Hale-Bopp Inferred from its
  Mid-Infrared Spectrum}. \apj 595, 522--530.

\bibitem[{{Pilipp} et~al.(1998){Pilipp}, {Hartquist}, {Morfill}, and
  {Levy}}]{1998A&A...331..121P}
{Pilipp}, W., {Hartquist}, T.~W., {Morfill}, G.~E., {Levy}, E.~H., Mar. 1998.
  {Chondrule formation by lightning in the Protosolar Nebula?} \aap 331,
  121--146.

\bibitem[{{Preibisch} et~al.(1993){Preibisch}, {Ossenkopf}, {Yorke}, and
  {Henning}}]{1993A&A...279..577P}
{Preibisch}, T., {Ossenkopf}, V., {Yorke}, H.~W., {Henning}, T., Nov. 1993.
  {The influence of ice-coated grains on protostellar spectra}. \aap 279,
  577--588.

\bibitem[{Servoin and Piriou(1973)}]{Servoin}
Servoin, J.~L., Piriou, B., 1973. Infrared reflectivity and {R}aman scattering
  of {M}g$_2${S}i{O}$_4$ single crystal. Phys. Stat. Sol. (b) 55, 677--686.

\bibitem[{{Spitzer} and {Kleinman}(1960)}]{1960PhRv..121.1324S}
{Spitzer}, W.~G., {Kleinman}, D.~A., 1960. {Infrared lattice bands of quartz}.
  Physical Review 121, 1324--1335.

\bibitem[{{Toon} and {Ackerman}(1981)}]{1981ApOpt..20.3657T}
{Toon}, O.~B., {Ackerman}, T.~P., Oct. 1981. {Algorithms for the calculation of
  scattering by stratified spheres}. \ao 20, 3657--3660.

\bibitem[{{van Boekel} et~al.(2004){van Boekel}, {Min}, {Leinert}, {Waters},
  {Richichi}, {Chesneau}, {Dominik}, {Jaffe}, {Dutrey}, {Graser}, {Henning},
  {de Jong}, {K{\" o}hler}, {de Koter}, {Lopez}, {Malbet}, {Morel}, {Paresce},
  {Perrin}, {Preibisch}, {Przygodda}, {Sch{\" o}ller}, and
  {Wittkowski}}]{2004Natur.432..479V}
{van Boekel}, R., {Min}, M., {Leinert}, C., {Waters}, L.~B.~F.~M., {Richichi},
  A., {Chesneau}, O., {Dominik}, C., {Jaffe}, W., {Dutrey}, A., {Graser}, U.,
  {Henning}, T., {de Jong}, J., {K{\" o}hler}, R., {de Koter}, A., {Lopez}, B.,
  {Malbet}, F., {Morel}, S., {Paresce}, F., {Perrin}, G., {Preibisch}, T.,
  {Przygodda}, F., {Sch{\" o}ller}, M., {Wittkowski}, M., Nov. 2004. {The
  building blocks of planets within the `terrestrial' region of protoplanetary
  disks}. \nat 432, 479--482.

\bibitem[{{van Boekel} et~al.(2005){van Boekel}, {Min}, {Waters}, {de Koter},
  {Dominik}, {van den Ancker}, and {Bouwman}}]{vanBoekelMin2005}
{van Boekel}, R., {Min}, M., {Waters}, L.~B.~F.~M., {de Koter}, A., {Dominik},
  C., {van den Ancker}, M.~E., {Bouwman}, J., 2005. {A 10$\,\mu$m spectroscopic
  survey of Herbig Ae star disks: grain growth and crystallization}. \aap,
  \emph{in press}.

\bibitem[{van~de Hulst(1957)}]{vandeHulst}
van~de Hulst, H.~C., 1957. Light Scattering by Small Particles. Wiley, New
  York.

\bibitem[{Warren et~al.(1994)Warren, Barret, Dodson, Watts, and
  Zolensky}]{DustCatalog}
Warren, J.~L., Barret, R.~A., Dodson, A.~L., Watts, L.~A., Zolensky, M.~E.
  (Eds.), 1994. Cosmic Dust Catalog. Vol.~14. {NASA} Johnson Space Center,
  Houston.

\bibitem[{{Waters} et~al.(1996){Waters}, {Molster}, {de Jong}, {Beintema},
  {Waelkens}, {Boogert}, {Boxhoorn}, {de Graauw}, {Drapatz}, {Feuchtgruber},
  {Genzel}, {Helmich}, {Heras}, {Huygen}, {Izumiura}, {Justtanont}, {Kester},
  {Kunze}, {Lahuis}, {Lamers}, {Leech}, {Loup}, {Lutz}, {Morris}, {Price},
  {Roelfsema}, {Salama}, {Schaeidt}, {Tielens}, {Trams}, {Valentijn},
  {Vandenbussche}, {van den Ancker}, {van Dishoeck}, {Van Winckel},
  {Wesselius}, and {Young}}]{1996A&A...315L.361W}
{Waters}, L.~B.~F.~M., {Molster}, F.~J., {de Jong}, T., {Beintema}, D.~A.,
  {Waelkens}, C., {Boogert}, A.~C.~A., {Boxhoorn}, D.~R., {de Graauw}, T.,
  {Drapatz}, S., {Feuchtgruber}, H., {Genzel}, R., {Helmich}, F.~P., {Heras},
  A.~M., {Huygen}, R., {Izumiura}, H., {Justtanont}, K., {Kester}, D.~J.~M.,
  {Kunze}, D., {Lahuis}, F., {Lamers}, H.~J.~G.~L.~M., {Leech}, K.~J., {Loup},
  C., {Lutz}, D., {Morris}, P.~W., {Price}, S.~D., {Roelfsema}, P.~R.,
  {Salama}, A., {Schaeidt}, S.~G., {Tielens}, A.~G.~G.~M., {Trams}, N.~R.,
  {Valentijn}, E.~A., {Vandenbussche}, B., {van den Ancker}, M.~E., {van
  Dishoeck}, E.~F., {Van Winckel}, H., {Wesselius}, P.~R., {Young}, E.~T., Nov.
  1996. {Mineralogy of oxygen-rich dust shells.} \aap 315, L361--L364.

\bibitem[{{Wehrstedt} and {Gail}(2002)}]{2002A&A...385..181W}
{Wehrstedt}, M., {Gail}, H.-P., Apr. 2002. {Radial mixing in protoplanetary
  accretion disks. II. Time dependent disk models with annealing and carbon
  combustion}. \aap 385, 181--204.

\bibitem[{{Weidenschilling}(1997)}]{1997Icar..127..290W}
{Weidenschilling}, S.~J., Jun. 1997. {The Origin of Comets in the Solar Nebula:
  A Unified Model}. Icarus 127, 290--306.

\bibitem[{{Weiler} et~al.(2003){Weiler}, {Rauer}, {Knollenberg}, {Jorda}, and
  {Helbert}}]{2003A&A...403..313W}
{Weiler}, M., {Rauer}, H., {Knollenberg}, J., {Jorda}, L., {Helbert}, J., May
  2003. {The dust activity of comet C/1995 O1 (Hale-Bopp) between 3 AU and 13
  AU from the Sun}. \aap 403, 313--322.

\bibitem[{{Wooden}(2002)}]{2002EM&P...89..247W}
{Wooden}, D.~H., 2002. {Comet Grains: Their IR Emission and Their Relation to
  ISm Grains}. Earth Moon and Planets 89, 247--287.

\bibitem[{{Wooden} et~al.(1999){Wooden}, {Harker}, {Woodward}, {Butner},
  {Koike}, {Witteborn}, and {McMurtry}}]{1999ApJ...517.1034W}
{Wooden}, D.~H., {Harker}, D.~E., {Woodward}, C.~E., {Butner}, H.~M., {Koike},
  C., {Witteborn}, F.~C., {McMurtry}, C.~W., Jun. 1999. {Silicate Mineralogy of
  the Dust in the Inner Coma of Comet C/1995 01 (Hale-Bopp) Pre- and
  Postperihelion}. \apj 517, 1034--1058.

\end{thebibliography}
\end{document}